\documentclass[%
 reprint,
 superscriptaddress,
 amsmath,amssymb,
 aps,
 pra,
]{revtex4-2}

\usepackage{graphicx}% Include figure files
\usepackage{dcolumn}% Align table columns on decimal point
\usepackage{bm}% bold math
\usepackage{hyperref}% add hypertext capabilities
\usepackage[dvipsnames]{xcolor}
\definecolor{mblue}{RGB}{42, 54, 144} % dark blue
\hypersetup{backref,pdfpagemode=FullScreen,colorlinks=true,breaklinks,urlcolor=mblue,linkcolor=mblue,citecolor=mblue}
\usepackage{placeins}
\usepackage{multirow}
\usepackage{siunitx}
\usepackage{braket}
\DeclareSIUnit\gauss{G}
\begin{document}

\preprint{APS/123-QED}

\title{Observation of low-field Feshbach resonances between $^{161}$Dy and $^{40}$K}

\author{Zhu-Xiong~Ye}
\affiliation{Institut f{\"u}r Experimentalphysik, Universit{\"a}t Innsbruck, Austria} 

\author{Alberto~Canali}
\affiliation{Institut f{\"u}r Experimentalphysik, Universit{\"a}t Innsbruck, Austria}

\author{Elisa~Soave}
\affiliation{Institut f{\"u}r Experimentalphysik, Universit{\"a}t Innsbruck, Austria} 

\author{Marian~Kreyer}
\affiliation{Institut f{\"u}r Experimentalphysik, Universit{\"a}t Innsbruck, Austria}

\author{Yaakov~Yudkin}
\affiliation{Institut f{\"u}r Experimentalphysik, Universit{\"a}t Innsbruck, Austria}
\affiliation{Department of Physics and QUEST Center, Bar-Ilan University, Ramat-Gan 5290002, Israel} 

\author{Cornelis~Ravensbergen}
\altaffiliation[Present address: ]
{ColdQuanta,Oxford Centre for Innovation, OX1 1BY, United Kingdom}
\affiliation{Institut f{\"u}r Experimentalphysik, Universit{\"a}t Innsbruck, Austria} 

\author{Emil~Kirilov}
\affiliation{Institut f{\"u}r Experimentalphysik, Universit{\"a}t Innsbruck, Austria} 
\affiliation{Institut f{\"u}r Quantenoptik und Quanteninformation (IQOQI), {\"O}sterreichische Akademie der Wissenschaften, Innsbruck, Austria}

\author{Rudolf~Grimm}
\affiliation{Institut f{\"u}r Experimentalphysik, Universit{\"a}t Innsbruck, Austria} 
\affiliation{Institut f{\"u}r Quantenoptik und Quanteninformation (IQOQI), {\"O}sterreichische Akademie der Wissenschaften, Innsbruck, Austria}

\date{\today}

\begin{abstract}

We report on the observation of Feshbach resonances at low magnetic field strength (below 10\,G) in the Fermi-Fermi mixture of $^{161}$Dy and $^{40}$K. We characterize five resonances by measurements of interspecies thermalization rates and molecular binding energies. As a case of particular interest for future experiments, we consider a resonance at 7.29\,G, which combines accurate magnetic tunability and access to the universal regime of interactions with experimental simplicity. We show that lifetimes of a few 100\,ms can be achieved for the optically trapped, resonantly interacting mixture. We also demonstrate the hydrodynamic expansion of the mixture in the strongly interacting regime and the formation of DyK Feshbach molecules. Our work opens up new experimental possibilities in view of mass-imbalanced superfluids and related phenomena.

\end{abstract}

\maketitle

\section{\label{sec:Introduction}Introduction}

Exciting research avenues in the field of quantum gases rely on the possibility to control interactions via magnetically tuned Feshbach resonances~\cite{Chin2010fri}. For ultracold systems composed of fermionic atoms~\cite{Inguscio2008ufg}, such resonances play a major role in pair formation and eventually in the creation of fermionic superfluids~\cite{Zwerger2012tbb,Strinati2018tbb,Bennemann2014both}. The majority of experiments on resonantly interacting fermions have been carried out on spin mixtures of fermionic species, which naturally imposes a limitation on equal-mass systems. Beyond that, theoretical work has predicted fermionic quantum gases with mass imbalance to favor exotic interaction regimes~\cite{Gubbels2013ifg}. Mass-imbalanced systems hold particular promise in view of superfluid states with unconventional pairing mechanisms~\cite{Gubbels2009lpi,Wang2017eeo,Pini2021bmf}, most notably the elusive Fulde-Ferrell-Larkin-Ovchinnikov (FFLO) state~\cite{Fulde1964sia,Larkin1964nss,Radzihovsky2010ifr}.

Ultracold heteronuclear Fermi-Fermi mixtures have so far been produced in the laboratory using four different combinations of fermionic atoms: $^6$Li-$^{40}$K~\cite{Taglieber2008qdt,Wille2008eau,Voigt2009uhf,Naik2011fri},
$^6$Li-$^{173}$Yb~\cite{Hara2011qdm,Green2020fri},
$^{161}$Dy-$^{40}$K~\cite{Ravensbergen2018poa,Ravensbergen2020rif}, and
$^{6}$Li-$^{53}$Cr~\cite{Neri2020roa,Ciamei2022euc,Ciamei2022DDF}.
Feshbach resonances have been found in all these combinations, but (with one exception~\cite{Ravensbergen2020rif}) they all turned out to be quite narrow. On the practical side, this limits the possibility of accurate experimental interaction tuning. On a more fundamental side, the narrowness imposes severe limitations to universal behavior in the resonance region~\cite{Chin2010fri} and thus also compromises the great benefit of fermionic loss suppression in a strongly interacting fermion mixture~\cite{Petrov2004wbd,Petrov2005dmi,Jag2016lof}.
The only exception is a rather broad resonance that we recently identified in the Dy-K mixture at a field of 217\,G~\cite{Ravensbergen2020rif}. Its experimental application, however, turned out to be challenging because of complications owing to the exceptionally rich structure of both Dy-K interspecies and Dy intraspecies resonances, which lead to a complex behavior of inelastic losses~\cite{Burdick2016lls,Soave2022lff}. This has motivated our search for alternative Feshbach resonances in the Dy-K system.

In this article, we report on the observation and characterization of five interspecies Feshbach resonances in the low-field region (below 10\,G). Although all of them belong to the class of narrow resonances~\cite{Chin2010fri}, they offer very interesting properties for experiments. Most notably, we find that a resonance near 7.29\,G offers the possibility of accurate magnetic tuning in combination with access to resonant interactions in the universal regime. In Sec.~\ref{sec:FRs}, we briefly introduce the basic properties of narrow Feshbach resonances. In Sec.~\ref{sec:servey}, we present a survey of the resonances in the low-field region, characterizing their strengths and the binding energies of the underlying molecular states. In Sec.~\ref{sec:7G}, we consider the particularly interesting 7.29-G resonance in more detail. We demonstrate that this low-field resonance offers an application potential similar to the 50 times wider resonance at 217\,G, but with less experimental complications. As examples, we present the hydrodynamic expansion of the mixture and the formation of DyK Feshbach molecules. In Sec.~\ref{sec:conclusion}, we finally summarize the main conclusions of our work.

\section{Narrow Feshbach resonances}\label{sec:FRs}

Near a single, isolated $s$-wave Feshbach resonance, elastic scattering is described by the scattering length $a$ as a function of the magnetic field strength $B$, following the expression~\cite{Moerdijk1995riu,Chin2010fri}
\begin{equation}
    \label{equ:FeshbachModel}
    a(B) = a_{\rm bg} - \frac{A}{B-B_0} \, a_0 \, .
\end{equation}
Accordingly, the near-resonant scattering behavior in the zero-energy limit is fully characterized by a set of three parameters: the resonance position $B_0$, the background scattering length $a_{\rm bg}$, and the strength parameter $A$. The latter is related to the common definition of a resonance width $\Delta$ by $A = \Delta \, a_{\rm bg}/a_0$ and provides a direct measure for the strength of the resonance pole. The Bohr radius $a_0$ is used as a convenient unit of length. 

To characterize a narrow (closed-channel dominated) Feshbach resonance~\cite{Chin2010fri}, an additional parameter is needed. Following Ref.~\cite{Petrov2004tbp} it is convenient to introduce the range parameter 
\begin{equation}
    \label{equ:Rstar}
    R^* = \frac{\hbar^2}{2 m_{\rm r} a_0 \, \delta \mu \, A}
\end{equation}
as an additional length scale, where $m_{\rm r}$ denotes the reduced mass and $\delta \mu$ represents the differential magnetic moment between the molecular state (closed channel) underlying the resonance and the atomic scattering state (entrance channel). Using this definition, the binding energy of the molecular state can be modeled as~\cite{Petrov2004tbp,Lous2018pti}
\begin{equation} 
\label{equ:bindingEnergyCurve}
E_{\rm b} = \frac{\hbar^2}{8 \, (R^*)^2 \, m_{\rm r}} \, \left( \sqrt{1- \frac{4 R^* (B-B_0)}{a_0 A}} -1 \right)^2 \, .
\end{equation}
In addition to that, $R^*$ also characterizes the effective range of scattering, corresponding to the momentum dependence at low, but finite collision energies.

For $|a| \gg R^*$, i.e.\ in a narrow range of magnetic detunings $|B-B_0| \ll A\,a_0/R^*$, one recovers universal behavior as it is the case for broad (entrance-channel dominated) resonances. The molecular binding energy then reduces to the universal expression $E_{\rm b} = \hbar^2/(2 m_{\rm r} a^2)$. An important experimental benefit of the universal regime is also the Pauli suppression of inelastic losses~\cite{Petrov2004wbd,Petrov2005dmi,Jag2016lof}.

\section{\label{sec:servey}Survey of low-field Feshbach resonances}

In this Section, we present a survey of the interspecies Feshbach resonances that we found in the magnetic field range from 0 to 10\,G. In Sec~\ref{subsec:prep}, we summarize the main experimental sequence for the preparation of the ultracold Dy and K mixture. In Sec.~\ref{sec:InterThemScan}, we introduce interspecies thermalization scans to detect the resonances and to estimate their relative strengths. In Sec.~\ref{sec:BEM}, we describe measurements of the binding energy for each resonance and their analysis by the model introduced in the preceding Section.

\subsection{\label{subsec:prep}Preparation of the mixture}

The starting point of our experiments is a Fermi-Fermi mixture of $^{161}$Dy and $^{40}$K atoms in a crossed-beam optical dipole trap (ODT), realized with near-infrared (1064\,nm) light. The preparation procedure is based on our previous work~\cite{Ravensbergen2018poa}, with implementation of an additional narrow-line (741 nm) in-trap cooling stage for Dy and further optimizations as described in detail in Appendix~\ref{app:preparation}.
Both species are spin polarized in their lowest hyperfine sublevels $\ket{F, m_F} = \ket{21/2, -21/2}$ and $\ket{9/2, -9/2}$, respectively. 

Under typical experimental conditions, forced evaporative cooling results in a mixture of $N_{\rm Dy} = 8\times 10^4$ and $N_{\rm K}=9 \times 10^3$ atoms in a trap with mean (geometrically averaged) oscillation frequencies of $\bar{\omega}_{\rm Dy}/2\pi = 150$\,Hz and $\bar{\omega}_{\rm K}/\bar{\omega}_{\rm Dy} = 3.60$~\cite{Ravensbergen2018ado} at a temperature of $T_{\rm Dy} = 70$\,nK. This corresponds to deeply degenerate conditions with $T_{\rm Dy}/ T_{F}^{\rm Dy}= 0.13$, 
where $T_{F}^{\rm Dy} = \hbar \bar{\omega}_{\rm Dy} (6 N_{\rm Dy})^{1/3} /k_{B}$ denotes the Dy Fermi temperature. The K temperature $T_{\rm K} = 90$\,nK is somewhat higher than for Dy because of a possibly incomplete interspecies thermalization in the final stage of the evaporation~\cite{Ravensbergen2018poa} but, in any case, we reach deep degeneracy also for the K component with $T_{\rm K} / T_{F}^{\rm K}\approx0.09$. The improvement with respect to our previous work~\cite{Ravensbergen2018poa} is essentially due to the additional narrow-line cooling stage for Dy described in Appendix~\ref{app:preparation}, which provides us with a lower initial temperature before evaporation. The final temperature of the sample can be controlled by variation of the endpoint of evaporation, and the number ratio $N_{\rm Dy}/N_{\rm K}$ by the initial conditions for loading the ODT from a laser-cooled sample.

While evaporative cooling is carried out at a magnetic field of 230\,mG, where we found three-body losses to be extremely low~\cite{Soave2022lff}, experiments near interspecies Feshbach resonances in general require a transfer of the sample to higher magnetic fields. The system then inevitably has to cross many interspecies Dy-K resonances and intraspecies Dy resonances, the latter exhibiting an extremely large density of narrow Feshbach resonances (typically 50 resonances per gauss~\cite{Soave2022lff}). To minimize loss and heating we decompress the sample by loading it adiabatically into a very shallow ODT with $\bar{\omega}_{\rm Dy}/2\pi = 39$\,Hz before we quickly (within a few ms) ramp up the magnetic field to its target value. 

\begin{figure*}[tb!]
\centering
\includegraphics[trim=10 30 15 0,clip,width=1.8\columnwidth]{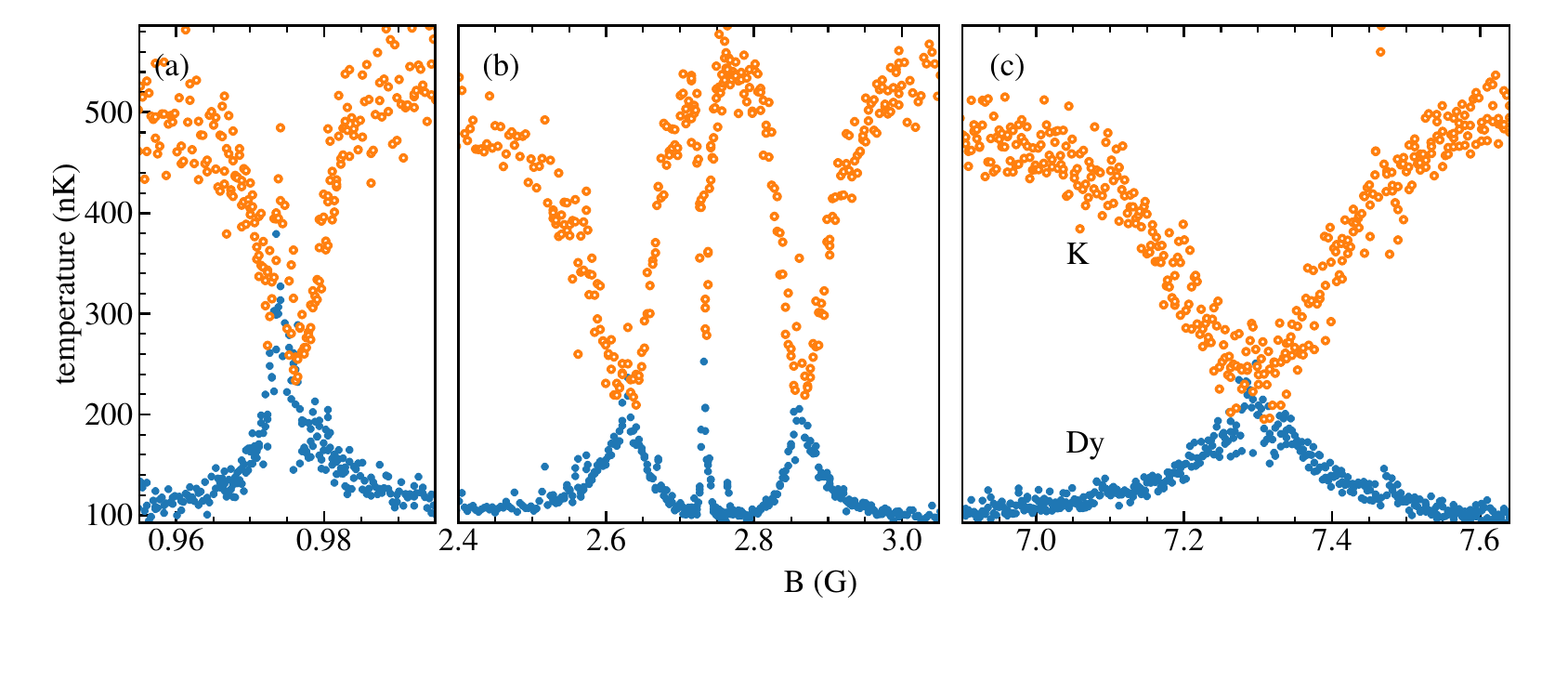}
\caption{Interspecies thermalization scans revealing five Feshbach resonances below 10\,G. The hold time is 50 ms for (a) and 100 ms for (b) and (c). Note the ten-fold extended $B$-field scale in (a) as compared to (b) and (c). Further note the narrow heating feature observed near 7.47\,G, which indicates an additional ultranarrow Feshbach resonance.}
\label{fig:ThermalizationLowerThan10G}
\end{figure*}

The shallow ODT is very sensitive to the effect of gravity, in particular for the heavy Dy atoms. Holding Dy in such a trap requires magnetic levitation~\cite{Weber2003bec,Ravensbergen2018poa}. A full compensation of the gravitational force for Dy is achieved by a gradient of 2.83\,G/cm. A gradient slightly below this value (2.69\,G/cm) leads to the `magic' levitation condition introduced in Ref.~\cite{Lous2017toa}, where both species experience the same gravitational sag and thus feature maximum overlap in the trap. A controlled deviation from this condition gives us a handle to vary the spatial overlap between the two species. This also allows us to separate both species, still keeping them in the trap. In this way, detrimental effects of interspecies resonances can be avoided during the transfer to the target magnetic field. This is of particular importance if many interspecies resonances need to be crossed to reach the final field, as is the case for experiments near the strong 217-G resonance~\cite{Ravensbergen2020rif}. Under optimized conditions, we can accomplish a transfer of a deeply cooled mixture from the evaporation field to target fields up to $\sim$10\,G without significant losses and with minimal heating to {$T_{\rm Dy}/ T_{F}^{\rm Dy}= 0.16$}. Note, however, that most of the experiments reported here are carried out at higher temperatures with moderately degenerate or near-degenerate mixtures.

At the end of the sequence, standard absorption images are taken to record the optical depth of each species after time-of-flight expansion. K images are taken after a few ms, while the expansion time for Dy is typically twice as long. The imaging for experiments in this paper is performed at a magnetic field of 230\,mG, except for the experiments on hydrodynamic expansion and Feshbach molecules (Secs.~\ref{sec:HydExp} and~\ref{sec:FRMole}), which are imaged at a magnetic field near 7\,G. For extracting the temperatures from the density profiles of deeply degenerate samples in the weakly interacting regime, we use fits based on the polylogarithmic function.

\subsection{\label{sec:InterThemScan}Interspecies thermalization scan}

Interspecies thermalization provides us with a powerful tool to investigate elastic scattering between the two species~\cite{Mosk2001mou}. In Ref.~\cite{Ravensbergen2020rif} we applied such measurements to characterize a scenario of broad overlapping Feshbach resonances in the $^{161}$Dy-$^{40}$K mixture at high magnetic fields (150\,-\,250~G) and to identify a particularly strong one centered near 217\,G. Here we proceed in an analogous way, focusing on the low-field region up to about 10\,G.

We interrupt the evaporation after cooling the mixture to 110\,nK with ${N_{\rm Dy}=3.3\times10^{4}}$ and $N_{\rm K}=1.8\times10^{4}$, and decompress the mixture to a low trap frequency of $\bar{\omega}_{\rm Dy}/2\pi=45.2$\,Hz and $\bar{\omega}_{\rm K}/2\pi=164$\,Hz. We apply species-selective parametric heating by modulating the intensity of the ODT at twice the vertical trapping frequency of K to induce a temperature difference between Dy and K. The temperature of K is increased to approximately 500\,nK within 100 ms of parametric heating, while the temperature of Dy remains constant. At this point, the mixture is near thermal with $T_{\rm Dy}/T_F^{\rm Dy}\approx0.9$ and $T_{\rm K}/T_F^{\rm K}\approx1.3$. Afterwards, we ramp the magnetic field rapidly (within 5\,ms) to the variable target field, where it is held for typically 100\,ms to allow for interspecies thermalization. The fixed hold time is chosen for optimum contrast in the thermalization spectrum. We switch the magnetic field back to 230\,mG within 5\,ms, before carrying out time-of-flight diagnostics. The temperatures and atom numbers of Dy and K are extracted from their density profiles after a 12-ms and a 5-ms time of flight, respectively.

We observe five interspecies resonances (along with a few ultranarrow, hardly resolved features) in the range from 0 to \SI{10}{G}. Figure\,\ref{fig:ThermalizationLowerThan10G}\,\cite{SM} shows the temperatures reached for Dy and K after the hold time versus magnetic field. Away from the resonances, the temperatures are very close to their initial values because of the relatively small interspecies cross section associated with the background scattering length $a_{\rm bg}$ (see also Sec.~\ref{sec:CSA}). Close to a resonance, the enhanced interspecies collisions cause Dy and K atoms to reach thermal equilibrium rapidly. We identify two isolated resonances at 0.97 and 7.29\,G, see Fig.\,\ref{fig:ThermalizationLowerThan10G} (a) and (c), and a group of three resonances at 2.61, 2.72, and 2.86\,G, see Fig.\,\ref{fig:ThermalizationLowerThan10G} (b).

The observed thermalization behavior is consistent with the one expected for Feshbach resonances in $s$-wave scattering. In particular, the observed five features do not disappear at very low temperatures, in contrast to higher partial wave resonances identified in the dense spectra of magnetic lanthanide atoms~\cite{Maier2015eoc, Khlebnikov2019rtc}. Our further analysis is therefore based on the assumption of an $s$-wave character and the model presented in Sec.~\ref{sec:FRs}.

The widths of the thermalization features contain information on the values of the strength parameter {$A=\Delta a_{\rm bg}/a_0$} as characterizing the different resonances according to Eq.\,(\ref{equ:FeshbachModel}). Following our thermalization model~\cite{Ravensbergen2020rif,Mosk2001mou}, the temperature difference decreases exponentially according to $\exp(-A^2Ct/(B-B_0)^2)$, where $t$ is the hold time at the target magnetic field and $C$ is a constant determined by the experimental parameters. Even without knowledge of the constant $C$, our features provide relative information on the values of $A$. Using the broadest feature (at 7.29\,G) for normalization, we obtain $A_{\rm rel}$= 0.06, 0.41, 0.05, 0.28, and 1 for our five resonances (in increasing field order). The relative $A$ parameters indicate that the Feshbach resonances at 2.61\,G and 2.86\,G have a similar strength, which is about 1/2 to 1/3 of the value at 7.29\,G. For the resonances at 0.97\,G and 2.72\,G, the $A$ parameter values are about 20 times smaller than that of 7.29\,G, showing that they are very narrow resonances. Compared with the much broader, previously observed resonance at 217\,G~\cite{Ravensbergen2020rif}, the value of $A$ is about 60 times smaller for the 7.29-G resonance.

Note that the magnetic-field scan in Fig.~\ref{fig:ThermalizationLowerThan10G}(a) reveals a narrow heating feature, which is located slightly below the resonance center. We also observe (not shown) substantial three-body recombination losses right at this point and we thus interpret the heating as a manifestation of an anti-evaporation effect~\cite{Weber2003tbr}, resulting from losses in the center of the trap. The corresponding scan of the much broader resonance in Fig.~\ref{fig:ThermalizationLowerThan10G}(c) does not show such a heating feature, but we find it to appear in measurements at longer hold times (see Sec.~\ref{sec:Lifetime}). We speculate that this behavior is a consequence of a stronger Pauli suppression of recombination losses for broader resonances.

In addition to the resonances presented here, we explore the Dy-K Feshbach resonances in the range of 10 to 20\,G and 50 to 60\,G. We observe two resonances located at 13.5\,G and 15.3\,G and 12 resonances distributed in the range of 51 to 55\,G, with two isolated resonances located at 51.2\,G and 54.9\,G, respectively. Based on these observations we can give a rough estimate of one resonance per 1.5\,G for the density of interspecies Feshbach resonances in the Dy-K system. This can be compared with a recent experimental investigation of Feshbach resonances in the Er-Li system~\cite{Schafer2022fro}, where about one resonance per 10\,G was observed consistent with theoretical predictions~\cite{Gonzalezmartinez2015mtf}. The higher density in the Dy-K system can be (at least partially) attributed to the smaller rovibrational spacing resulting from the about six times larger reduced mass.

\subsection{\label{sec:BEM}Binding energy measurements}

To further characterize the observed $s$-wave Feshbach resonances, we measure the binding energy of the DyK dimer at the different resonances by wiggling the magnetic field~\cite{Chin2010fri,Claussen2003vhp}. The mixture is prepared in the decompressed ODT with mean trap frequencies of $\bar{\omega}_{\rm Dy}=2\pi\times36.8$ Hz and $\bar{\omega}_{\rm K}=2\pi\times132$\,Hz in thermal equilibrium with a temperature of $T\approx370$\,nK. To obtain a strong signal, the atom number ratio between Dy and K is prepared to be around 1:1 ($N_{\rm Dy}\approx N_{\rm K}\approx3\times10^4$), and the magic levitation field is applied to ensure maximum overlap. The magnetic field is modulated in a frequency range between 10\,kHz and 3\,MHz. If the modulation frequency matches the binding energy of the DyK dimer at the selected magnetic field, the free atom pairs are associated into weakly bound molecules, which causes losses due to inelastic collisions. We adjust the modulation amplitude and duration to produce a clearly visible loss feature, where the loss of atoms is between 1/3 and 2/3 of the initial value. The typical magnetic field modulation amplitude and duration are roughly 400\,mG and 200\,ms, respectively. The loss features show typical widths of 20 to 40 kHz (full width at half maximum), and we fit them with a Lorentzian curve to determine the resonant frequency. We calibrate the magnetic field by measuring the frequency of the K transition $|F,m_F\rangle =|9/2,-9/2\rangle \rightarrow |9/2,-7/2\rangle$ as described in Appendix~\ref{app:BCali}.

Figure~\ref{fig:BindingEnergyCurve} displays the binding energy versus magnetic field, measured near 1\,G, 3\,G, and 7\,G. The solid curves show the best fits to the data according to Eq.\,(\ref{equ:bindingEnergyCurve}), with $B_0$, $A$, and $\delta \mu$ as free parameters, and $R^*$ is calculated from $A$ and $\delta \mu$ using Eq.\,(\ref{equ:Rstar}). The resulting parameter values are summarized in {Table~\ref{tab:ParamtersOfFBRs}}. In addition, we have carried out binding energy measurements (data not shown) for the isolated resonance near 51\,G and analyzed them in the same way. The results are included in Table~\ref{tab:ParamtersOfFBRs}. 

\begin{figure*}[t]
\centering
\includegraphics[trim=0 5 10 0,clip,width=1.8\columnwidth]{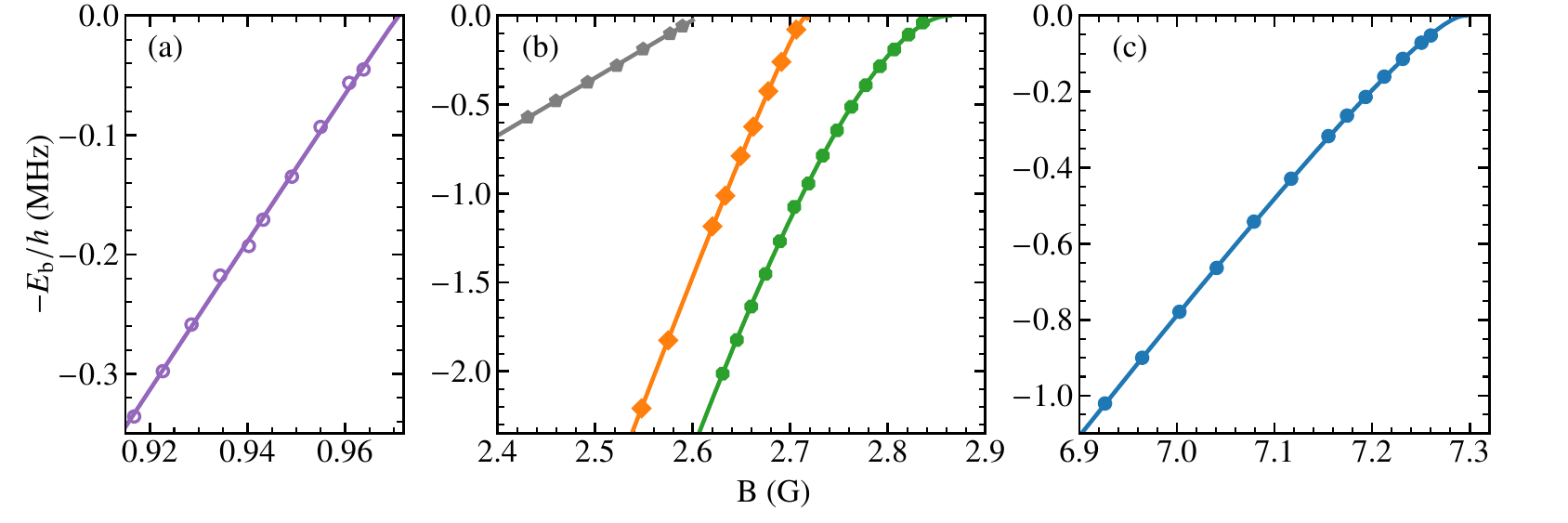}
\caption[.]{Binding energy of Feshbach molecules related to the observed resonances below 10\,G. The symbols represent experimental data measured by wiggling the magnetic field, and the solid lines are the fitted curves based on Eq.\,(\ref{equ:bindingEnergyCurve}). The uncertainties are smaller than the size of symbols. Note the five-fold extended $B$-field scale in (a) as compared to (b) and (c).}
\label{fig:BindingEnergyCurve}
\end{figure*}

\begin{table}[b]
\caption{\label{tab:ParamtersOfFBRs}
Parameters of the observed Feshbach resonances from fits of {Eq.\,(\ref{equ:bindingEnergyCurve})} to the measured binding energies. The uncertainties represent the 1\,$\sigma$ statistical errors from the fit. They do not include additional uncertainties (2\,mG for $B<10$\,G, and {10\,mG} for $B\approx50$\,G) from the magnetic field calibration, see Appendix~\ref{app:BCali}. The symbol $\mu_B$ represents the Bohr magneton. Note that the three resonances near 3\,G are presumably too close to each other to justify an interpretation in terms of isolated resonances, which obscures the physical meaning of the fit parameters (see text).}
\begin{ruledtabular}
\begin{tabular}{cccccc}
$B_0$ (G)& $A$ (G)& $\delta\mu/\mu_B$ & $R^*/a_0$\\
 \hline
0.971(2) &0.6(2.3) &4.5(3)  &$\gtrsim3000$\\
2.608(2) &0.0(1.8) &2.3(1)  &$\gtrsim10^4$\\
2.717(1) &4.4(3) &11.1(1)   &831(70)\\
2.864(1) &28.1(7) &22(1)     &65(5)\\
7.295(1) &23.2(9) &2.71(3)   &643(30)\\
\hline
{51.242(2)}  &{15(1)}   &4.2(1) &646(60)\\
\end{tabular}
\end{ruledtabular}
\end{table}

It is important to note that the model underlying Eq.\,(\ref{equ:bindingEnergyCurve}) is based on the assumption of a single, isolated resonance, not overlapping with other resonances. This is the case for the resonances at 0.97, 7.29, and 51.24\,G, but not for the group of three resonances near 3\,G. Even though the data can be well fitted by Eq.\,(\ref{equ:bindingEnergyCurve}), the fit parameters will lose their direct physical meaning and they will no longer reflect the real physics. In particular, the value extracted for the strength parameter $A$ will no longer correspond to the value of $A$ that describes the behavior of the scattering length near the pole according to Eq.\,(\ref{equ:FeshbachModel}). In a similar way, the value extracted for the asymptotic differential magnetic moment $\delta \mu$ may no longer reflect the magnetic moment of the closed-channel molecule, and the true pole position may also be somewhat shifted from the best fit value of $B_0$. In fact, comparing the fitted $A$ values for the overlapping resonances near 3\,G with each other and with the ones of the 7.29-G resonance shows that their relative values do not correspond to the behavior observed in the thermalization measurements. Also, the anomalously large fit value for the differential magnetic moment of the 2.86-G resonance appears likely as an artefact resulting from coupling between the different closed channels of the overlapping resonances. A proper description would require a more sophisticated coupled-channels model~\cite{Yudkin2021esf,Li2022erp}, which is beyond the scope of the present article.

The resonance at 7.29\,G has no significant overlap with other resonances, and we can therefore expect Eqs.~(\ref{equ:FeshbachModel}-\ref{equ:bindingEnergyCurve}) to describe the situation properly. The resulting value of the strength parameter (${A\approx }$ 23\,G) highlights that a magnetic detuning of $|B-B_0| =$\,10\,mG already leads to a large value for the scattering length of $|a|\approx$ 2400$a_0$. The universal range of the resonance (see the bending of the binding energy curve in Fig.\,\ref{fig:BindingEnergyCurve}(c) near threshold) is realized for $|B-B_0|\ll A a_0 / R^* =$ 36\,mG. In experiments at low magnetic fields~\cite{Soave2022lff}, control of the field strength to a few mG is rather straightforward, and the resonance thus appears to be well suited for exploring universal physics at large values of the scattering length. Note that the resonance near 51\,G offers very similar properties, but at a higher magnetic field.

\section{\label{sec:7G}Case study of the 7.29-G resonance}

The 7.29-G resonance is the broadest Feshbach resonance with the largest universal range that we found in the low-field range up to 10\,G. Moreover, it is well isolated from other resonances, which allows for a description in terms of a two-channel model as underlying Sec.~\ref{sec:FRs}. Even in the range of up to 300\,G, it is one of very few resonances offering similar properties. The low magnetic field facilitates accurate experimental control, and it allows us to minimize atom loss and heating, which usually occurs when the field is ramped to its target value and other resonances need to be crossed. Because of these favorable properties, we examine this particular resonance more closely.

In Sec.~\ref{sec:CSA}, we start with presenting thermalization measurements at different magnetic field strengths, from which we obtain the relationship between scattering length and magnetic field. In Sec.\,\ref{sec:Lightshift}, we investigate a possible light shift of the resonance by measuring binding energies at different ODT depths. In Sec.\,\ref{sec:HydExp}, we present the hydrodynamic expansion of the strongly interacting mixture and investigate its resonance behavior by comparing our experimental results with Monte-Carlo simulation. In Sec.\,\ref{sec:Lifetime}, we study inelastic losses in the vicinity of resonance and provide an estimate for the three-body loss rate at the pole position. In Sec.~\ref{sec:FRMole}, we finally present the first results on the observation of DyK Feshbach molecules.

\subsection{\label{sec:CSA} Elastic scattering cross section and scattering length}

To determine the scattering length as a function of the magnetic field in the vicinity of the 7.29-G resonance, we measure the interspecies cross section between Dy and K atoms at different magnetic field strengths. The method is the same as we have already applied to the Dy-K mixture in our previous work~\cite{Ravensbergen2020rif,Ravensbergen2018poa}, and is based on the model introduced in Ref.~\cite{Mosk2001mou}. The model describes how the interspecies scattering cross section $\sigma$ can be extracted from the evolution of the temperature difference $\Delta T$ between Dy and K atoms.

\begin{figure}[tb]
%centering
\includegraphics[trim=0 10 0 0,clip,width=1\columnwidth]{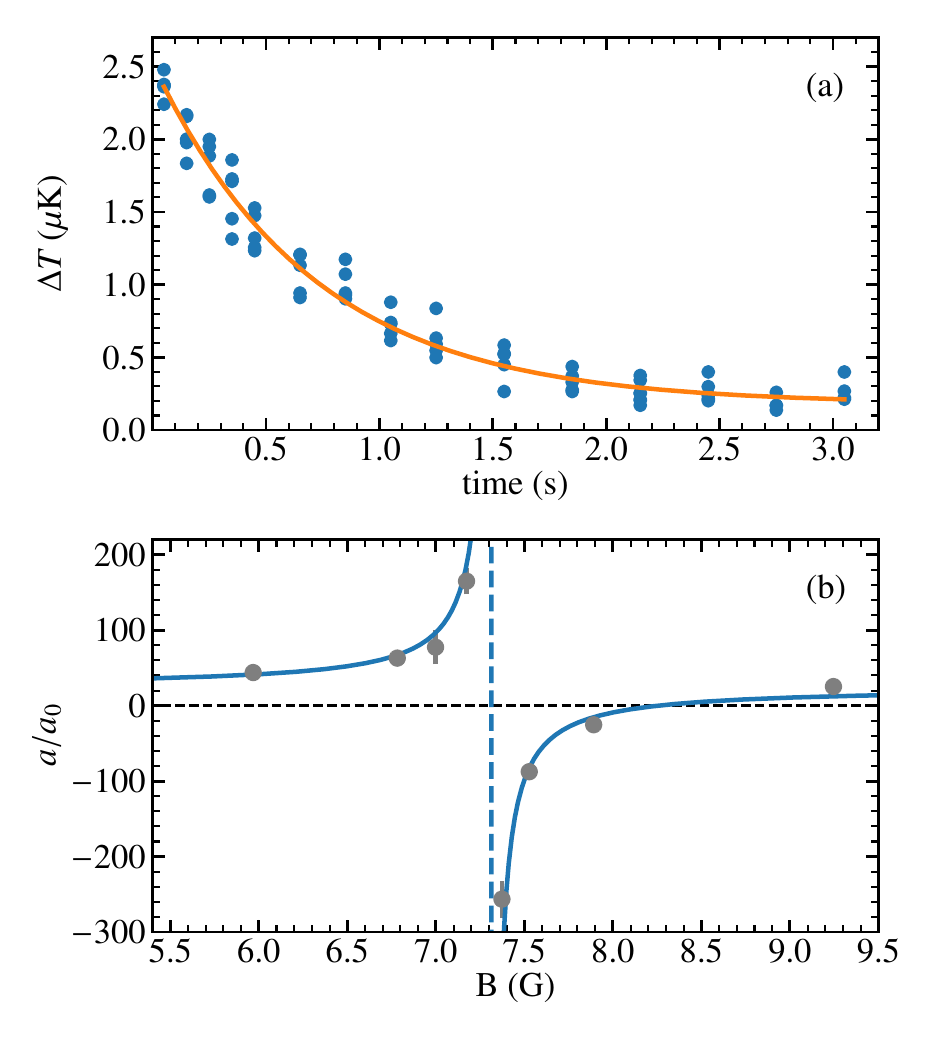}
\caption{\label{fig:TempEvoAndScatvsB}Scattering length measurement in the vicinity of the 7.29-G resonance.~(a) Temperature difference $\Delta T= T_{\rm K} - T_{\rm Dy}$ versus hold time in ODT at 5.97\,G. The orange solid line shows the numerical fit curve with the $s$-wave scattering length $a=53(9)a_0$.~(b) Scattering length between Dy and K versus magnetic field strength. The gray points represent the values measured by interspecies thermalization. The solid blue line is a fit based on Eq.\,(\ref{equ:FeshbachModel}), and the dashed vertical line indicates the pole position.}
\end{figure}

As an example, Fig.\,\ref{fig:TempEvoAndScatvsB}(a) illustrates a typical thermalization process at 5.97\,G. Here, the initial atom numbers are $N_{\rm Dy} =8\times10^4$ and $N_{\rm K}=1.7\times10^4$, and the mixture is prepared in a trap with $\bar{\omega}_{\rm Dy}=2\pi\times88$\,Hz and $\bar{\omega}_{\rm K}=2\pi\times318$\,Hz. After species-selective heating, see Sec.~\ref{sec:InterThemScan}, the initial temperatures are ${T_{\rm Dy}=\SI{0.5}{\mu K}}$ and ${ T_{\rm K}=\SI{2.5}{\mu K}}$. Both species approach thermal equilibrium on the time scale of 1\,s. The orange solid line shows a numerical fit, which yields an interspecies scattering length of magnitude $|a|=53(9)a_0$. 

The gray dots in Fig.\,\ref{fig:TempEvoAndScatvsB}(b) show the values of the scattering length measured with the same method at different magnetic field strengths~\cite{BChoice}. The corresponding signs are determined by the position of the magnetic field relative to the resonance pole. As a general fact, the sign of the scattering length is positive on the low-field side of a Feshbach resonance in the lowest spin channel. In this channel, the atom-pair state carries the maximum magnetic moment and thus always experiences a larger Zeeman shift than the molecular state underlying the Feshbach resonance.

The solid blue line shows the best fit based on Eq.\,(\ref{equ:FeshbachModel}). The parameter values show a background scattering length of $a_{\rm bg}=+23(5){a_0}$, which is smaller than earlier results of $|a_{\rm bg}|\approx 62a_0$ obtained at 430\,mG~\cite{Ravensbergen2018poa} and $|a_{\rm bg}|\approx 60a_0$ in the 200-G region~\cite{Ravensbergen2020rif}. The resonance width according to the common definition~\cite{Chin2010fri} is $\Delta = Aa_0/a_{\rm bg}$. For the width $\Delta$ and the position of the zero crossing $B_{\rm zc} = B_0 + \Delta$, we obtain $\Delta=0.94(21)$\,G and $B_{\rm zc} = 8.25(20)$\,G from a fit with variable $B_0$, and $\Delta=1.02(21)$\,G and $B_{\rm zc} = 8.32(21)$\,G from a fit with $B_0$ fixed to 7.925\,G, the value that was obtained from the binding energy measurements.

As shown in Table~\ref{tab:ParametersAt7G}, thermalization and binding energy measurements give consistent results on the resonance position $B_0$ and the resonance strength $A$. This also supports the validity of our assumption of an isolated Feshbach resonance, where the binding energy follows Eq.\,(\ref{equ:bindingEnergyCurve}).

\begin{table}[t]
\caption{\label{tab:ParametersAt7G} Comparison of fit parameter values related to the 7.29-G Feshbach resonance extracted from different observations (see text).}
\begin{ruledtabular}
\begin{tabular}{ccccccc}
Method &Section&$B_0$ (G)& $A$ (G) &$R^*/a_0$ &$a_{\rm bg}/a_0$\\
 \hline
Binding energy&\ref{sec:BEM}&7.295(1)  &23.2(9) &643(30) &-\\
\multicolumn{1}{c}{\multirow{2}{*}{Thermalization}} & \multicolumn{1}{c}{\multirow{2}{*}{\ref{sec:CSA}}}&{7.314(20)} &{22.8(2.6)}  &-&{24.2(4.7)}\\
\multicolumn{1}{c}{}& \multicolumn{1}{c}{}&{7.295\footnote{The pole position $B_0$ is fixed to the value obtained from the binding energy measurements.}} &{23.4(2.4)}  &-  &{22.8(4.5)}\\
Hydrodynamics&\ref{sec:HydExp}&7.290(2) &23.2\footnote{The strength parameter $A$ is fixed to the value obtained from the binding energy measurements.}&-&-\\
\end{tabular}
\end{ruledtabular}
\end{table}

\subsection{\label{sec:Lightshift}Light shift}

A differential polarizability between the entrance channel (atom pair state) and the closed channel (molecular state) in the infrared optical trap may cause a light-induced shift in the position of a Feshbach resonance. We have observed such an effect in the $^6$Li-$^{40}$K Fermi-Fermi mixture~\cite{Jag2014ooa} and the $^6$Li-$^{41}$K Fermi-Bose mixture~\cite{Lous2018pti}. Taking this shift into account turned out to be essential for accurate interaction tuning, and also proved to be a very useful tool to implement extremely fast changes of the interaction strength~\cite{Cetina2016umb}. Here we investigate whether the same effect is present in our Dy-K mixture.

\begin{figure}[t]
\includegraphics[trim=10 10 10 10,clip,width=0.8\columnwidth]{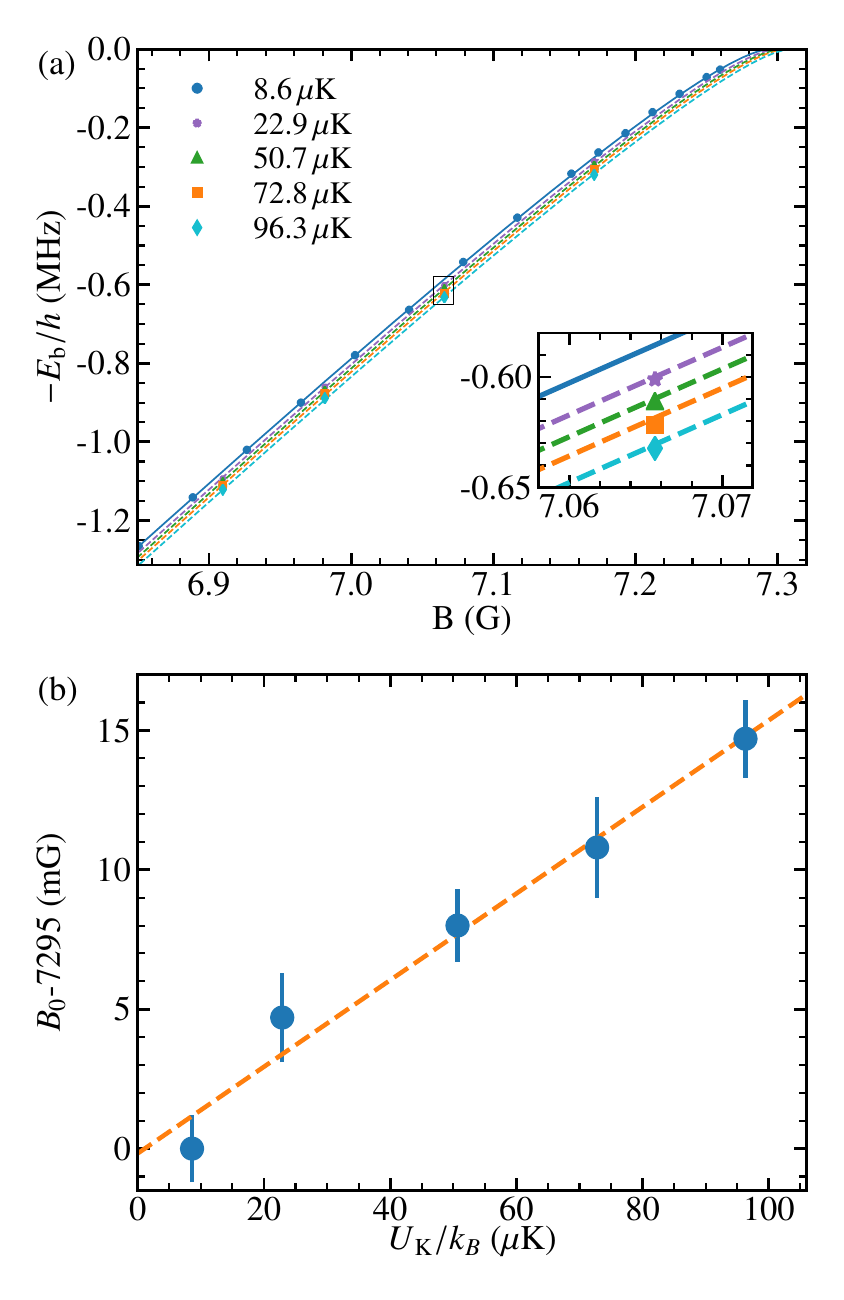}
\caption{\label{fig:lightshift}Light shift of the Feshbach resonance. (a) Binding energy versus magnetic field for different ODT depths. The symbols represent the measured data, and the solid and dashed lines show the fits based on Eq.\,(\ref{equ:bindingEnergyCurve}), from which we extract the resonance position $B_0$. The solid blue curve is the same as shown in Fig.\,\ref{fig:BindingEnergyCurve}(c). (b) Resonance position $B_0$ versus the trap depth for the K atoms. The blue bullets show the fitted results of (a). The orange dashed line represents the linear fit according to ${B_0=\xi U_{K}+B_{\rm zero}}$, with $\xi$=$1.6(5)\times10^{-4}$\,\si{G/\mu K} and ${B_{\rm zero}=7.295(3)}$\,G.}
\end{figure}

To characterize this effect on the 7.29-G resonance, the Dy-K mixture is prepared in traps of different depths, then the same approach as described in Sec.~\ref{sec:BEM} is applied to measure the binding energy at different magnetic field strengths. Fig.\,\ref{fig:lightshift}(a) shows our results with the solid blue line representing our reference measurement as shown in Fig.\,\ref{fig:BindingEnergyCurve}(c). The inset of Fig.\,\ref{fig:lightshift}(a) shows a zoom-in of the binding energy near 7.065\,G. We observe a small frequency shift of the order of a few kHz between adjacent points. The dashed curves represent the fits to the data points based on Eq.\,(\ref{equ:bindingEnergyCurve}) with only $B_0$ as a free parameter, keeping $A$ and $\delta \mu$ fixed to the values of the reference curve. The best fit values for the pole position are shown as blue circles in Fig.\,\ref{fig:lightshift}(b). The orange dashed line represents the linear fit ${B_0=\xi U_{\rm K}+B_{\rm zero}}$, which yields the pole position without light shift, ${B_{\rm zero}=7.295(3)}$\,G, and the linear slope $\xi=1.6(5)\times10^{-4}$\,\si{G/\mu K}. This value for the slope $\xi$ is about one order of magnitude smaller than typical values observed in Li-K mixtures~\cite{Lous2018pti,Jag2014ooa}. Such a small slope implies that the light shift effect is negligible for the shallow traps used in our experiment. It also tells us that the optical trapping potential for Feshbach molecules at 7.29\,G can be simply regarded as the sum of the trap depths for free Dy and K atoms with only $3\%$ deviation~\cite{MoltrapDepth}.
 
\subsection{\label{sec:HydExp}Hydrodynamic expansion}

In Ref.~\cite{Ravensbergen2020rif}, we reported on hydrodynamic expansion as a striking effect of resonant interspecies interaction. On top of the broad 217-G Feshbach resonance, the two components were observed to expand jointly in a collisionally dense cloud after release from the trap. A Monte-Carlo simulation of the resonant collisional dynamics showed excellent agreement with the experimental data. Here we carry out essentially the same experiment, but on the 7.29-G resonance, which is more than 50 times weaker than the 217-G resonance.

The sequence here is similar to the one applied in the experiments described before, but without the decompression stage at the end of evaporation. The mixture is held in a tight ODT with the magic levitation field, where the trap frequencies are $\bar{\omega}_{\rm Dy}/2\pi=134$\,\si{Hz} and $\bar{\omega}_{\rm K}/2\pi=484$\,\si{Hz}. To minimize atom loss and heating, the magnetic field is rapidly (in 2\,ms) ramped from the evaporation field to 7.42\,G and held for 50\,ms to ensure a stable field strength. Here, the long lifetime of the Dy-K mixture and relatively large scattering length of $a\approx-200a_0$ allow Dy and K atoms to reach thermal equilibrium in a short time without inelastic losses and heating. {At this particular field, also Dy losses~\cite{Soave2022lff} are found to be very weak.} We get $N_{\rm Dy}=6.5\times10^4$ and $N_{\rm K}=2\times10^4$ atoms with a temperature of $T=\SI{740}{nK}$, corresponding to $T_{\rm Dy}/T_F^{\rm Dy}=1.6$ and $T_{\rm K}/T_F^{\rm K}=0.6$. These conditions are close to the ones of Ref.~\cite{Ravensbergen2020rif}. To monitor the hydrodynamic expansion, the magnetic field is ramped to its target value in 2\,ms and held for a further 2\,ms in the ODT to establish a stable $B$ field. After that, the ODT and the levitation field are switched off simultaneously, and time-of-flight absorption imaging is applied to record the optical depth profile of Dy and K atoms at the target magnetic field.

The resonant interaction regime is realized when the $s$-wave scattering length exceeds all other relevant length scales. These are the inverse wave number of relative motion $1/\bar{k}_{\rm rel}\approx1700a_0$, the inverse Fermi momentum $1/k_{F}\approx1370a_0$, and the range parameter $R^*=643a_0$. For the present experiments, we obtain a range of $\pm$\SI{15}{mG}, about 50 times narrower than on the broad 217-G resonance explored in previous experiments.

\begin{figure}[tb]
\includegraphics[trim=15 10 15 10,clip,width=1\columnwidth]{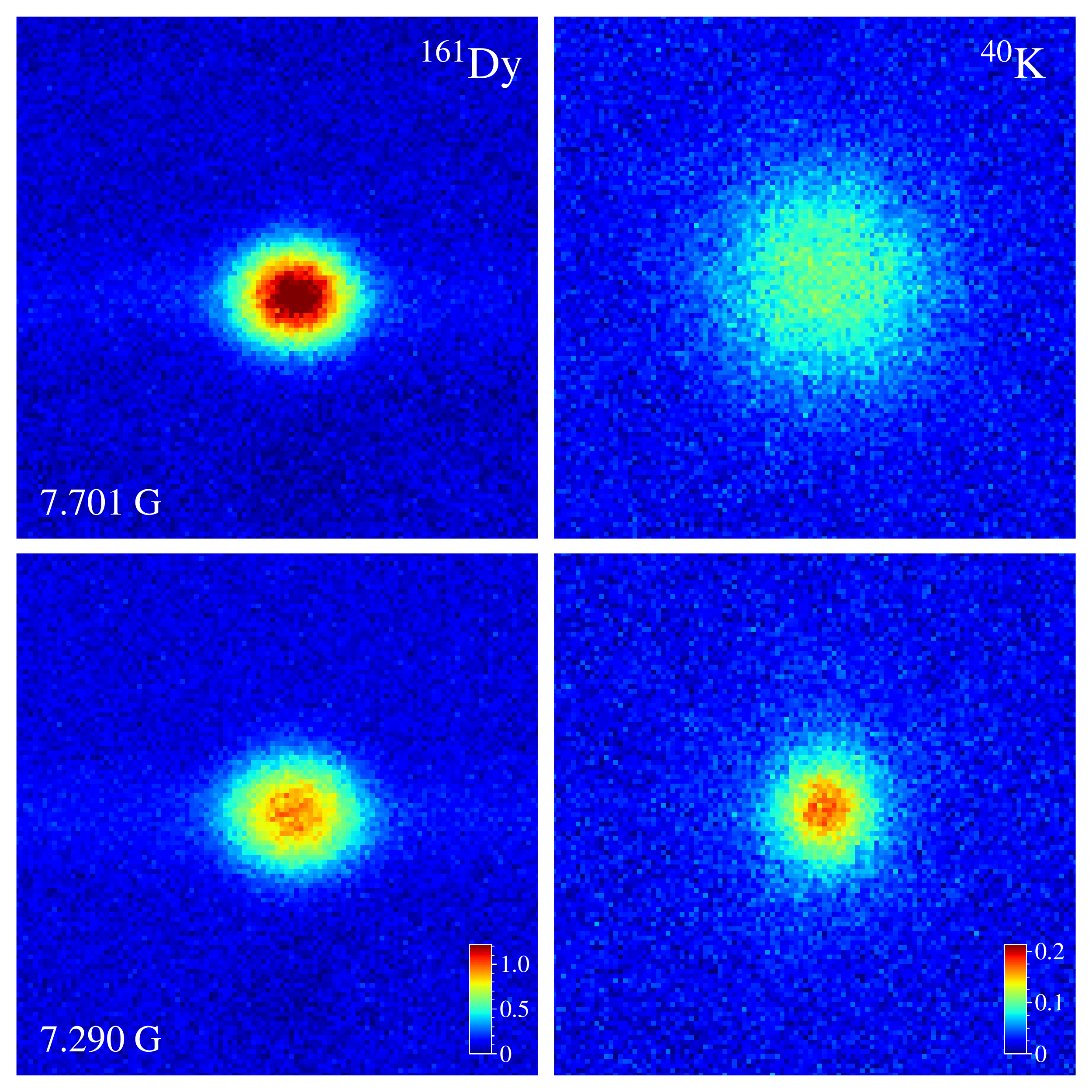}
\caption{\label{fig:Hyrodynamic} Absorption images of Dy and K far away from (upper row) and on (lower row) the 7.29-G Feshbach resonance after 4.5\,ms of free expansion. The optical depth of K is significantly enhanced on resonance, while Dy decreases because of the collective expansion. The field of view of all images is \SI{360}{\mu m}$\times$\SI{360}{\mu m}. Each image is an average of five individual shots.}
\end{figure}

In Figure.\,\ref{fig:Hyrodynamic}, we demonstrate the observed hydrodynamic behavior in the time-of-flight expansion of the mixture. The upper row shows the optical depths of Dy atoms and K atoms at \SI{7.701}{G} with the estimated scattering length $a\approx-38a_0$. Here, we do not observe any significant effect of interaction between the two species. The lighter potassium atoms expand faster than the heavier dysprosium atoms, leading to a size ratio of $\sigma_{\rm K}/\sigma_{\rm Dy}=\sqrt{m_{\rm Dy}/m_{\rm K}}\approx2$, where the $\sigma_{\rm K}$ and $\sigma_{\rm Dy}$ are defined as a standard deviation of the spatial distribution. The lower row in Fig.\,\ref{fig:Hyrodynamic} shows the expansion on resonance, where the strong interspecies collision rate leads to collective behavior. The size of the K cloud becomes much smaller than at 7.701\,G and its peak density is greatly enhanced. On the contrary, the size of the Dy cloud is slightly larger than at 7.701\,G with a decreased peak density. For both species, the clouds have a similar size on resonance because of the collective expansion effect, and a few K atoms can be observed outside the overlap region of Dy and K cloud. The latter are the atoms that expand freely after having escaped from the hydrodynamic core~\cite{Ravensbergen2020rif}.

\begin{figure}[t]
\includegraphics[trim=10 10 10 10,clip,width=0.9\columnwidth]{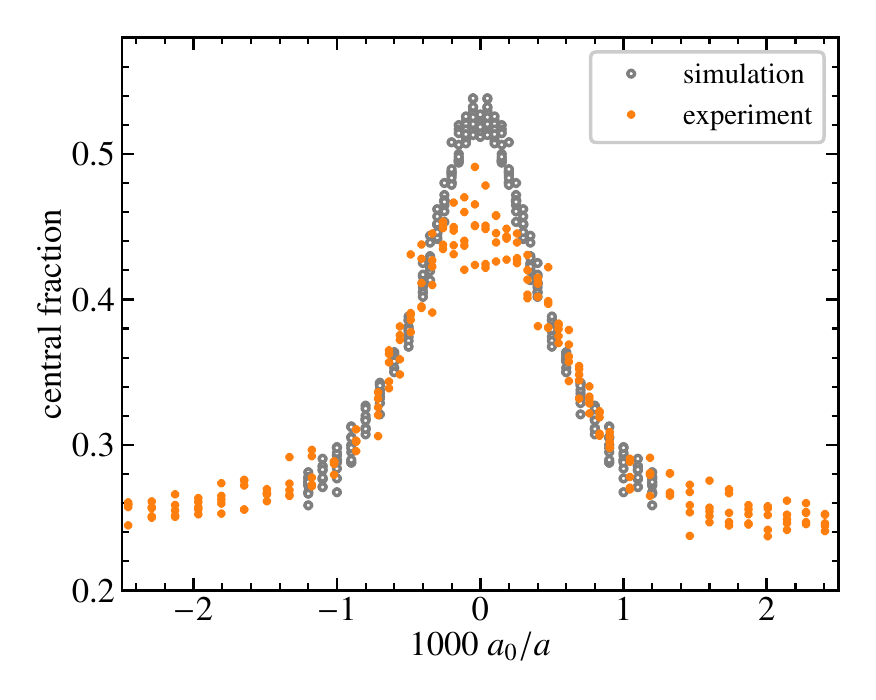}
\caption{\label{fig:HydroSimAndExp} Comparison of the central fraction as a function of scattering length obtained from Monte-Carlo simulations (gray open circles) and experiment (orange solid circles). The central fraction is defined as the fraction of K atoms inside $\sqrt{2}\sigma_{\rm Dy}$ region.}
\end{figure}

To quantitatively analyze the hydrodynamic expansion~\cite{Ravensbergen2020rif}, we explore the central fraction defined as the fraction of K atoms inside a circle with radius $\sqrt{2}\sigma_{\rm Dy}$. The orange dots in Fig.\,\ref{fig:HydroSimAndExp} show the experimental central fraction as a function of scattering length, which we calculated from Eq.\,(\ref{equ:FeshbachModel}) with a fixed $A=23.2$\,G (see Table~\ref{tab:ParamtersOfFBRs}). Far away from the resonance (at 7.701\,G), we get a central fraction of {0.22} (out of the range of Fig.\,\ref{fig:HydroSimAndExp}) corresponding to the expansion of two non-interacting Gaussian clouds. The central fraction gradually increases to 0.45 as the scattering length increases. The width at half maximum of experimental data is about $|1000a_0/a|=0.68$ from Lorentzian curve fit, which is consistent with our length scale estimated range of $B-B_0\approx\pm15$\,mG.

We also perform the same Monte-Carlo simulation as we did for the 217-G resonance~\cite{Ravensbergen2020rif} to extract the central fraction. We take into account the momentum dependence of the resonant collisional cross section, but neglect the small effect of $R^*$. The gray open circles in Fig.~\ref{fig:HydroSimAndExp} show our simulation results as a function of the inverse scattering length. We find that the matching between simulation and experiment is very sensitive to the exact value of the pole position $B_0$. Therefore hydrodynamic expansion allows us to determine the best value of $B_0$ = 7.290\,G with an uncertainty of 2\,mG, which is dominated by magnetic field fluctuations in our system.

As shown in Fig.~\ref{fig:HydroSimAndExp}, the experiment matches well with the simulation in most regions apart from a narrow central part. The good match confirms our value for the strength parameter $A$ obtained from the binding energy and the thermalization measurements (see Table\,\ref{tab:ParametersAt7G}). The value of the pole position here is slightly below the value derived from the binding energy measurements, which we attribute to a systematic error in the binding energy measurements caused by neglecting the small, but finite collision energies in the mixture. 
This indeed leads to a small overestimation of the binding energies and the fit produces a pole position, which is a few mG too high. Therefore, for the further discussion, we choose $B_0=7.290(2)$\,G as the presumably more accurate pole position.

We notice that the observed central fraction is smaller than the one from the simulation in the resonance range with $|a|\gtrsim3000a_0$, while we did not observe such a deviation in our previous experiments near the broad 217-G resonance~\cite{Ravensbergen2020rif}. We suspect that uncontrolled magnetic field fluctuations of the order of a few mG may have been caused by turning off the levitation gradient when the mixture is released from the trap, which may have reduced the effect of interactions very close to resonance during the free expansion. Obviously, the 7.29-G resonance is much more sensitive to magnetic field fluctuations than the broad 217-G resonance. Apart from this rather technical issue, we observe essentially the same behavior of the hydrodynamic expansion for both resonances.

We finally note that the expanding clouds in Fig.~\ref{fig:Hyrodynamic} show small anisotropies, which may be related to hydrodynamic behavior and/or the dipole-dipole interacting in the strongly magnetic Dy component. The origin of this effect requires further investigations using more anisotropic trapping schemes than applied in the present work and will be subject of future work.

\subsection{\label{sec:Lifetime}Lifetime and three-body decay}

Collisional stability is an important ingredient for deep evaporative cooling of the Dy-K mixture and for future experiments on mass-imbalanced fermionic superfluids~\cite{Gubbels2013ifg,Wang2017eeo,Pini2021bmf}. In Ref.~\cite{Ravensbergen2020rif}, we have reported long lifetimes of the mixture near the 217-G resonance, which we attributed to Pauli suppression of inelastic few-body processes~\cite{Petrov2004wbd,Petrov2005dmi}. Here we study whether we can achieve also comparably long lifetimes near the 7.29-G resonance.

\begin{figure}[t]
\includegraphics[trim=10 10 0 5,clip,width=1\columnwidth]{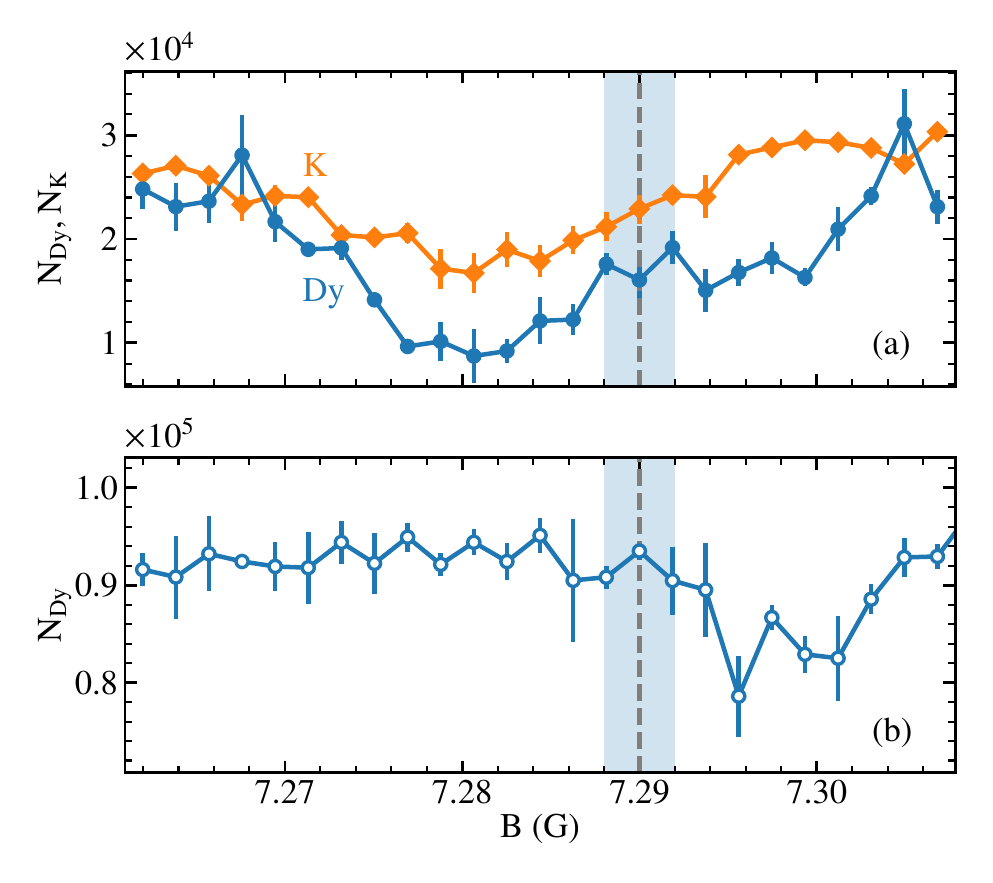}
\caption{\label{fig:LossfeatureAt7G}Loss scans in the vicinity of the 7.29-G resonance. (a) Atom number loss spectra of the two components of the mixture for a hold time of 140\,ms in the ODT. The initial atom numbers are $N_{\rm Dy}=3.4\times10^4$ and $N_{\rm K}=3.3\times10^4$ with a temperature of {$T=170$\,nK}. A pronounced interspecies loss feature appear about 10\,mG below resonance, where the $s$-wave scattering length is very large and positive. (b) Number of atoms in a pure Dy cloud in the same trap after a hold time of 220\,ms, with an initial atom number of {$N_{\rm Dy}\approx9.5\times10^4$} and a temperature of $T=140$\,nK. The gray dashed line indicates the pole position of the Feshbach resonance according to the hydrodynamic expansion measurements in Sec.~\ref{sec:HydExp}. The shaded region indicates the uncertainty. Without K, significant losses of Dy only show up 5-10\,mG above the resonance, while a 30-mG wide region below resonance appears to be essentially free of Dy loss features.}
\end{figure}

\begin{figure}[t]
\includegraphics[trim=0 5 10 0,clip,width=1\columnwidth]{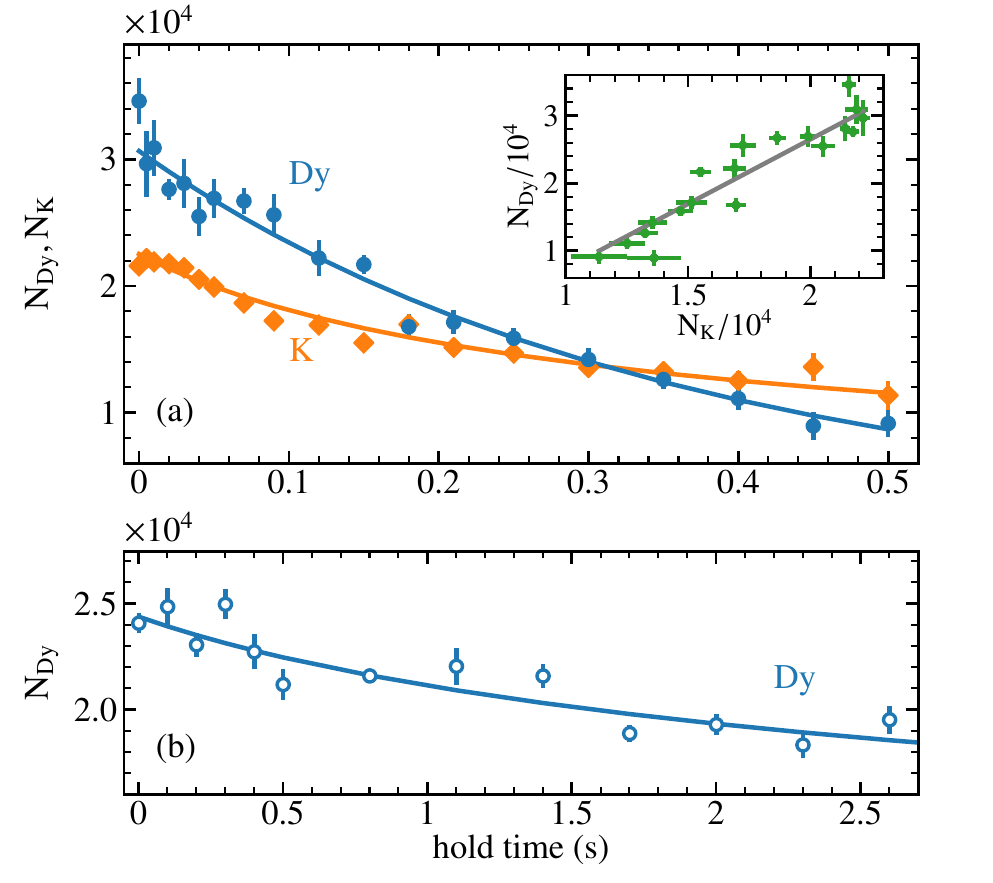}
\caption{\label{fig:lifetimeat7G}Comparison of atom number decay between (a) the Dy-K mixture and (b) a pure sample of Dy in the ODT at the pole position. The temperature of the mixture is $115$\,nK and of the pure Dy cloud is 49\,nK. The solid lines represent fits with the initial lifetime of $\tau_{\rm DyK}=360$\,ms for the mixture and $\tau_{\rm Dy}\approx5$\,s for Dy alone. The inset in (a) shows the relationship between the Dy and K atom numbers during the decay. The solid line represents a linear fit with $N_{\rm Dy}=aN_{\rm_K}+b$, yielding $a=1.9(2)$ and $b=-5.8(1.6)\times10^3$.}
\end{figure}

To initiate the measurement, the sample is prepared in a shallow ODT with a mean trap frequency of $\bar{\omega}_{\rm Dy}=2\pi\times35$\,Hz and $\bar{\omega}_{\rm K}=2\pi\times128$ Hz, where the initial temperature and atom number of the sample varies slightly for different measurements (see Figs.~\ref{fig:LossfeatureAt7G} and \ref{fig:lifetimeat7G}). The magnetic field is ramped to the target value and the atom numbers are recorded by absorption imaging after holding the mixture in the trap for {a few 100 ms}. 

In Fig.\,\ref{fig:LossfeatureAt7G}(a), we present the loss spectrum for the mixture in a 50-mG wide range, showing both the number of remaining Dy and K atoms. For comparison, in Fig.\,\ref{fig:LossfeatureAt7G}(b), we show the corresponding spectrum for a pure Dy sample, which is expected to show three-body loss features because of the generally very dense spectrum of resonances in fermionic Dy~\cite{Burdick2016lls,Soave2022lff}. The gray dashed line indicates the pole position of the resonance as obtained from the hydrodynamic expansion measurement presented in the preceding section. Figure\,\ref{fig:LossfeatureAt7G}(a) shows a striking ``valley'' for both species, located on the lower side of the resonance. This is about 10\,mG below the pole position, where we estimate a large positive value of the scattering length of $a\approx +2300a_0$. The loss feature appears in both Dy and K and is stronger in Dy by about a factor of two. A similar feature was observed for the 217-G resonance~\cite{Ravensbergen2020rif}, which also resembles observations made in experiments on resonantly interacting spin mixtures of fermions~\cite{Regal2004lom,Bourdel2003mot,Dieckmann2002doa}. We interpret this loss as being due to the formation of weakly bound dimers by three-body recombination and subsequent inelastic atom-dimer collisions. The fact that maximum losses are observed in the regime of weakly bound dimers and not on top of the resonance (as usually observed in bosonic systems) points to the presence of Pauli suppression on resonance. 

Regarding Dy intraspecies losses, Fig.\,\ref{fig:LossfeatureAt7G}(b) shows a wide plateau of negligible intraspecies losses along with a narrow resonant loss feature on the upper side of the resonance. We find that, for the Feshbach resonance in the Dy-K mixture, the pole position and the range of weakly bound universal dimers are located within the low-loss plateau, which is a lucky coincidence and very favorable for future experiments on the resonantly interacting mixture.

Figure~\ref{fig:lifetimeat7G}(a) shows the decay of the trapped mixture at 7.290\,G, i.e.\ at the very center of the resonance. The initial atom numbers are $N_{\rm Dy}=3.4\times10^4$ and $N_{\rm K}=2.2\times10^4$ with an initial temperature {$T=\SI{115}{nK}$}. We analyze the loss curve based on the assumption that three-body processes dominate the loss. Quantitatively, we extract the initial slope from a heuristic fit model as described in previous work~\cite{Ravensbergen2020rif,Soave2022lff}, which then allows us to derive a value for the three-body event rate coefficient. The solid curves in Figs.\,\ref{fig:lifetimeat7G}(a) and (b) represent the fitted curves. For the lifetime of the mixture, we obtain $\tau_{\rm DyK}=360$\,ms, which is much shorter than the one of Dy alone ($\tau_{\rm Dy}\approx5$\,s~\cite{DyTau}). Obviously, interspecies processes are the main cause of losses on resonance.

To identify the dominant loss process, we plot in the inset of Fig.\,\ref{fig:lifetimeat7G}(a) the number of Dy atoms versus the number of K atoms during the decay. A linear fit yields a slope of $dN_{\rm Dy}/dN_{\rm K} = 1.9(2)$, which suggests a three-body process involving two heavy and one light atom as the dominant one. This may be interpreted as a consequence of the effective three-body attraction in a resonant heavy-heavy-light fermion system~\cite{Levinsen2011ada, Jag2014ooa}. Neglecting other processes, the loss of atoms in the mixture can be modeled by the loss-rate equation
\begin{equation}
\begin{aligned}
\frac{d N_{\rm Dy}}{d t}=2\frac{d N_{\rm K}}{d t}=&-2K_3 \int n_{\rm Dy}^{2} n_{\rm K} d^{3} r,
\end{aligned}
\label{equ:K3coefficiency}
\end{equation}
where $n_{\rm Dy}$ and $n_{\rm K}$ are the position-dependent atomic number densities of Dy and K atoms, respectively. The event rate coefficient $K_3$ quantifies Dy-Dy-K three-body losses. Under our conditions, background losses are negligible.

To obtain an approximate value for the event rate coefficient $K_3$, we assume that the atomic number densities follow thermal Gaussian distributions. For Dy with $T/T_F^{\rm Dy} = 1.2$, the distribution is essentially thermal. The K component with $T/T_F^{\rm K}\approx 0.3$ is degenerate, so that Fermi pressure increases the cloud size. However, because of the tighter optical confinement of K, this has only little effect on the overlap integral in Eq.~(\ref{equ:K3coefficiency}). We also neglect any interaction effect on the spatial distributions. Under these assumptions, Eq.\,(\ref{equ:K3coefficiency}) leads to~\cite{Ravensbergen2020rif}
\begin{equation}
\frac{\dot{N}_{\mathrm{K}}}{N_{\mathrm{K}}}=-K_3 N^2_{\mathrm{Dy}}\left(\frac{\sigma}{2 \pi \sigma_{\mathrm{Dy}}^{2} \sigma_{\mathrm{K}}}\right)^{3}\,,
\label{equ:K3coefficiencySimplied}
\end{equation}
where $\sigma=(2\sigma^{-2}_{\rm Dy}+\sigma^{-2}_{\rm K})^{-1/2}$ and $\sigma_{i}=\omega^{-1}_i\sqrt{k_BT/m_i}$ ($i=$Dy, K). We finally obtain a value of $K_3=1.8(7)\times10^{-24}$\,\si{cm^6/s} for the three-body event rate coefficient on resonance.

We can compare the resulting $K_3$-value with measurements near the 217-G resonance in Dy-K mixture, related work on resonant Fermi-Bose mixtures of K-Rb and Li-K~\cite{Bloom2013tou, Lous2018pti}, and Bose-Bose mixtures of K and Rb~\cite{Wacker2016utb}. This indicates a Pauli suppression of inelastic few-body losses by roughly one order of magnitude, but not much more. In this respect, the 7.29-G resonance seems to be less favorable than the 217-G resonance. Nevertheless, with the very shallow ODT applied, we could demonstrate lifetimes of a few 100\,ms, which gives enough time to carry out experiments in the resonant regime. 

\subsection{\label{sec:FRMole}Creation of Feshbach molecules}

An important application of Feshbach resonances is the association of weakly bound dimers by magnetic field ramps~\cite{Chin2010fri}. A common way to detect such Feshbach molecules is Stern-Gerlach separation~\cite{Regal2003cum,Herbig2003poa}, facilitated by the different magnetic moment to mass ratio of dimers and free atoms. We here apply this technique to demonstrate the formation of weakly bound DyK molecules.

The starting point of these experiments is a degenerate mixture of $N_{\rm Dy} =6.4\times10^4$ and $N_{\rm K}=2.9\times10^4$ atoms in an ODT with $\bar{\omega}_{\rm Dy} =2\pi\times40.3$\,Hz and $\bar{\omega}_{\rm K} =2\pi\times146$\,Hz. The temperature of $T =125$\,nK corresponds to $T/T_F^{\rm Dy} =0.9$ and $T/T_F^{\rm K} =0.33$. The magic levitation gradient of 2.69\,G/cm is applied to eliminate the differential gravitational sag, see Sec.~\ref{subsec:prep}. The magnetic field is then ramped up to 7.62\,G, which is about 330\,mG above the resonance center. This particular field has been chosen to exploit a minimum of loss and heating related to Dy intraspecies processes. The ramp to this preparation field takes only 2\,ms, but additional 20\,ms are given for the magnetic field to settle to a stable value. 

The Feshbach molecules are created by sweeping the magnetic field to 7.11\,G within 2\,ms, i.e.\ at a rate of $\dot{B}=$\SI{-0.25}{G/ms}. At the end of the sweep, the ODT is switched off and the magnetic field gradient is raised to \SI{7.5}{G/cm} in \SI{0.5}{ms}. For the Dy atoms, this increased gradient results in a total, upward-directed acceleration of 1.65$g$ (2.65$g$ from the magnetic gradient and -$g$ from the field of gravity). For the DyK dimer (molecular magnetic moment of about $8.9\mu_B$), the acceleration amounts to 0.89$g$. The K cloud is nearly levitated with a small acceleration of only 0.07$g$. During a time of flight of 5\,ms, the DyK dimers are efficiently separated from free Dy and free K atoms, which puts the molecular cloud between the two atomic clouds. Finally, for detection, molecules are dissociated to free atoms by ramping back the magnetic field to \SI{7.62}{G} in \SI{1}{ms}, and absorption images are taken for Dy (K) atoms after further 2\,ms (1\,ms).

\begin{figure}[t]
\includegraphics[trim=0 0 0 0,clip,width=0.7\columnwidth]{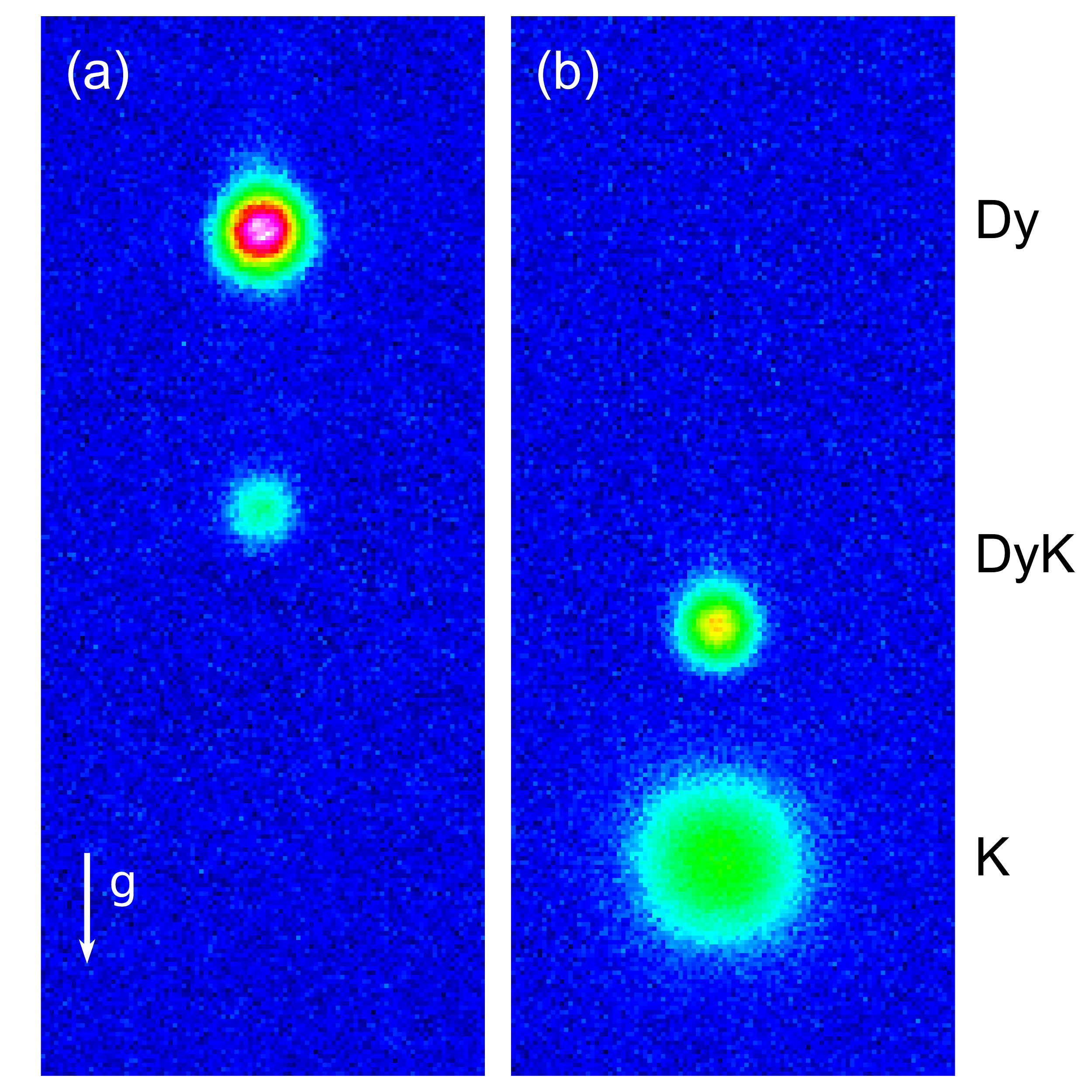}
\caption{\label{fig:moleculesAt7G}Stern-Gerlach analysis of the Dy-K mixture after association of dimers by a Feshbach ramp and dissociation by an inverse ramp. Absorption images of (a) Dy and (b) K are taken at the magnetic field of 7.62\,G with a gradient of 7.5\,G/cm. {The images in (a) and (b) are finally taken after a further 2\,ms and 1\,ms, after completion of the dissociation ramp.} The arrow shows the direction of gravity. The field of view of all images is \SI{329}{\mu m}$\times$\SI{785}{\mu m}.}
\end{figure}

In Fig.\,\ref{fig:moleculesAt7G}, we present the resulting absorption images of Dy atoms (a) and K atoms (b). The free Dy and K atomic clouds are at the top and bottom of the images, respectively. The smaller clouds located closer to the center of the images result from the dissociation of Feshbach molecules. Note that the Dy and K clouds resulting from the dissociating dimers are not exactly at the same vertical position. This is due to the different acceleration in the short time between dissociation and imaging. More than $N_{\rm mol}=7.5\times10^3$ molecules were created with a transfer efficiency of of $N_{\rm mol}/N_{\rm K,0} \approx 20\%$, where $N_{\rm K,0}$ is the K atom number right before the association ramp.

The small size of the expanding molecular cloud, as clearly seen in the images in comparison with the atomic clouds, points to a rather low temperature. From the temperature in expansion and the trap potential, we can infer a phase-space density of about 0.1 for the trapped molecular cloud~\cite{Lam2022hps}. With further optimization, this preliminary observation is very promising for reaching the quantum-degenerate regime, i.e.\ Bose-Einstein condensation of these molecules. Experiments pursuing this goal are underway in our laboratory.

\section{Summary and conclusion} \label{sec:conclusion}

In the Fermi-Fermi mixture of $^{161}$Dy and $^{40}$K, prepared in the lowest spin states, we have identified in total five Feshbach resonances in the magnetic-field region below 10\,G. We have characterized these resonances by observing the enhanced interspecies thermalization and by measuring binding energies of the molecular states underlying the resonances. We have extracted the main resonance parameters: position, strength, and magnetic moment of the molecular state. The five resonances can all be classified as narrow (closed-channel dominated) resonances~\cite{Chin2010fri}, but for some of them the universal range, where the physics of broad resonances is recovered, is wide enough to be conveniently addressable in experiments.

As a case of particular interest, we have investigated the resonance at 7.29\,G in more detail and demonstrated several applications. This resonance is well separated from other ones and can thus be modeled within a basic two-channel approach~\cite{Petrov2004tbp}. We have extracted the resonance strength in three different ways: Measured binding energies and elastic scattering in thermalization and in hydrodynamic expansion yielded consistent results, which confirms our good understanding of the resonance.

We have also studied the decay properties of the Dy-K mixture near the 7.29-G resonance and found a qualitative behavior very similar to our previous work~\cite{Ravensbergen2020rif} near the broad 217-G resonance in the same system. A loss maximum is found in the universal range on the lower side of the resonance, where the scattering length is large and positive and weakly bound molecules are produced by three-body recombination. Such a behavior is typical for fermionic mixtures with loss suppression on resonance.
Our measurement of three-body decay on top of the resonance reveals a relatively large decay rate in comparison with the 217-G resonance, which is in accordance with the general trend that broader resonances are more favorable for fermionic loss suppression.

Our experiments show that, employing the 7.29-G resonance, one can do essentially everything that has been done for the broad 217-G resonance. The low field offers major practical advantages in simplicity, speed, and accuracy of magnetic field control. Another very important advantage consists in the fact that the interspecies resonance region is not contaminated by the many intraspecies Dy resonances. A price one has to pay for these advantages seems to be a weaker suppression of inelastic decay, but our experiments have shown that, in a very shallow optical trap, lifetimes of a few hundred milliseconds can be achieved for the resonant mixture, which is enough for many experimental applications.

We have also presented a proof-of-principle demonstration of the creation of Feshbach molecules by sweeping across the resonance, which indicates a high efficiency of the process. Our near-future work will be dedicated to identifying the optimum conditions for molecule creation, the investigation of collisional processes involving the molecules, and to reaching the necessary phase-space densities for molecular Bose-Einstein condensation. This will be an important step towards our general goal to realize novel superfluid states in fermionic mixtures of ultracold atoms.

We finally note that the spectrum of interspecies Feshbach resonances in mixtures of submerged-shell lanthanide atoms with alkali atoms can be expected to be non-chaotic~\cite{Gonzalezmartinez2015mtf}, in contrast to intraspecies resonances in experiments with Dy and Er~\cite{Maier2015eoc}. Therefore, we are confident that, with input from our experiments, a scattering model can be developed that fully describes the resonance scenario and allows experimentalists to identify the best suited resonances for specific purposes.

\begin{acknowledgments}

We thank M. Zaccanti and D. Petrov for stimulating discussions and V. Corre and J. H. Han for contributions in the early stage of this work. The project has received funding from the European Research Council (ERC) under the European Union’s Horizon 2020 research and innovation programme (grant agreement No.\ 101020438 - SuperCoolMix). We further acknowledge support by the Austrian Science Fund (FWF) within Project P34104-N, and within the Doktoratskolleg ALM (W1259-N27). We also acknowledge support by the Erasmus+ Programme of the European Union and the Austria-Israel Academic Network Innsbruck (AIANI) at the University of Innsbruck. We thank the members of the ultracold atom groups in Innsbruck for many stimulating discussions and for sharing technological know-how.

\end{acknowledgments}

\appendix

\section{Preparation} \label{app:preparation}

The sources of Dy and K, the magneto-optical traps for the two species, the gray-molasses cooling stage for K, and the sequential dipole trap loading scheme have been described already in Ref.~\cite{Ravensbergen2018poa}. Here we discuss a more refined sequence of optical dipole trapping (Sec.~\ref{sec:appA}), which also facilitates in-trap narrow-line Doppler cooling (Sec.~\ref{sec:appB}). With these improvements and some further optimizations, we find that the atom numbers in the deeply degenerate regime of the mixture can be increased by typically a factor of five.

\subsection{Optical dipole trapping schemes}\label{sec:appA}

\begin{figure}[t]
\centering
\includegraphics[width=1\columnwidth]{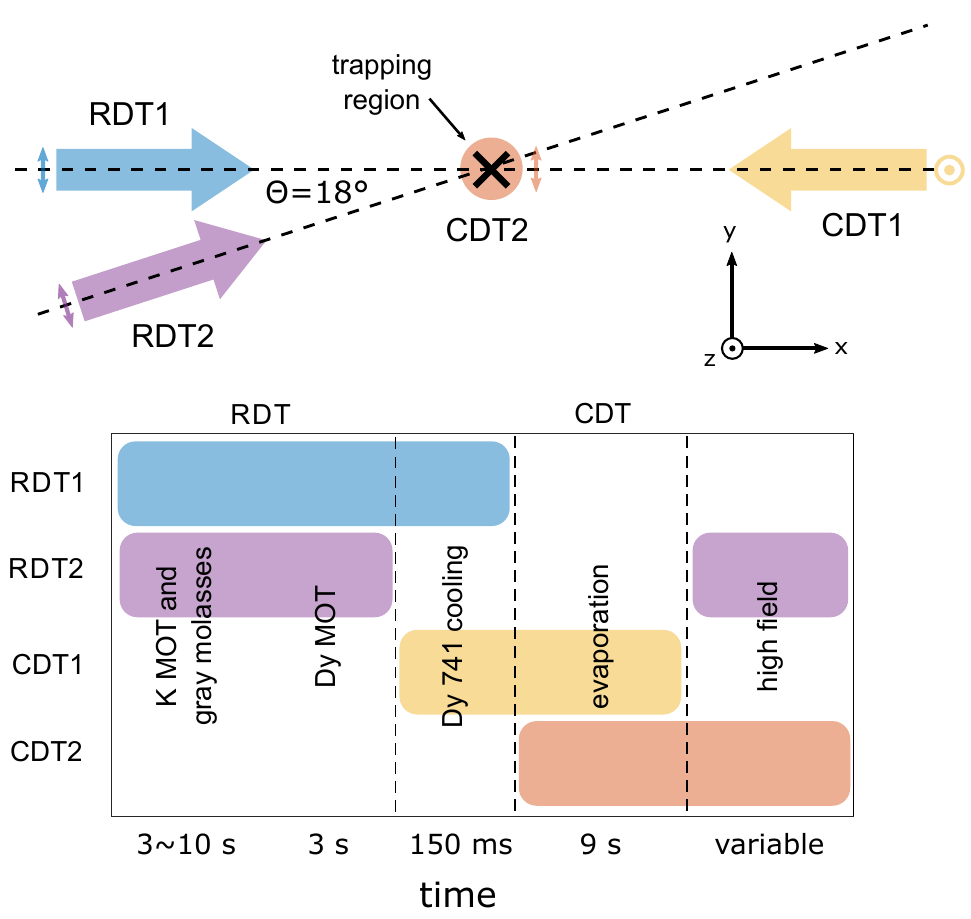}
\caption{Laser beam configuration for optical dipole trapping (top), and timing sequence (bottom) of the different combinations in the four trapping stages. While the large-volume reservoir dipole trap (RDT) is applied for loading, the tightly confining crossed dipole trap (CDT) is used for evaporative cooling. In an intermediate stage between RDT and CDT, in-trap narrow-line cooling is applied to Dy. In the last stage, the sample is decompressed and transferred to the desired target magnetic field, at which the specific experiments are carried out.}
\label{fig:odtscheme}
\end{figure}

Our preparation sequence relies on four stages of optical dipole trapping. We employ in total four infrared laser beams in the configuration illustrated in Fig.~\ref{fig:odtscheme}\,(top); the beam parameters are listed in Table~\ref{tab:odtbeams}. The timing scheme in Fig.~\ref{fig:odtscheme}\,(bottom) shows which combination of crossing beams is employed to realize the optical dipole traps in the four different stages.

\begin{table}[b]
\caption{\label{tab:odtbeams}
Trapping beam properties. Waist $w$ and maximum power $P_{\rm max}$ for each of the four beams. The two RDT (CDT) beams are derived from an Azurlight ALS-IR-1064-5-I-CC-SF (Mephisto MOPA 18 NE) and frequency offset by 160\,MHz (220\,MHz) by acousto-optical modulation.}
\begin{ruledtabular}
\begin{tabular}{lcccc}
\textrm{}&
\textrm{RDT1}&
\textrm{RDT2}&
\textrm{CDT1}&
\textrm{CDT2}\\
\colrule
$w$ ($\mu$m) & 94 & 93 & 24 & 62\\
$P_{\rm max}$ (W) & 12 & 11 & 0.4 & 0.73\\
\end{tabular}
\end{ruledtabular}
\end{table}

In the first stage, a large-volume reservoir dipole trap (RDT) facilitates efficient loading of both species from magneto-optical traps. This dipole trap is realized with two high-power infrared beams (labeled RDT1 and RDT2), crossing in the horizontal plane at an angle of $18^\circ$. In a later stage, in which evaporative cooling is performed, we use a crossed-beam dipole trap (CDT) realized with two tightly focused beams. The beam CDT1 propagates horizontally in opposite direction to RDT1, and CDT2 is directed vertically. These two dipole traps (RDT and CDT) for loading and evaporative cooling have been introduced in our previous work~\cite{Ravensbergen2018poa}. For the present work, we have introduced two upgrades:

(1) Between the RDT and the CDT stage, we have implemented an additional stage for narrow-line Doppler cooling of Dy. For this purpose, an elongated trap is realized with beams RDT1 and CDT1.

(2) After the CDT stage, the evaporatively cooled mixture is transferred into a shallow dipole trap formed by beams RDT2 and CDT2. This decompresses the sample, facilitates transfer to high magnetic fields with minimized collisional losses, and is used for most of the experiments reported here.

\subsection{In-trap narrow-line laser cooling of Dy}\label{sec:appB}

Laser cooling on the narrow 741-nm line of Dy~\cite{Lu2011soa} is carried out as an intermediate step before evaporative cooling. The line is an inner-shell transition ($4f\rightarrow5d$) with a narrow linewidth of 1.8\,kHz. The Doppler temperature of 43\,nK is an order of magnitude lower than the recoil temperature of 213\,nK, which makes it very interesting for deep laser cooling.

We generate the 741-nm light with an external-cavity diode laser (Moglabs ECDL) locked to a high-finesse cavity. The narrowing of the laser linewidth is achieved by a fast feedback loop acting directly on the diode current (Toptica mFALC 110). We combine the 741-nm cooling light with the 626-nm beams of the magneto-optical trap, using the same polarizations. A magnetic bias field of 230\,mG defines the quantization axis. The saturation intensity of the transition is as low as $0.57\,\mu$W/cm$^2$.

Doppler cooling is carried out in an elongated trap, into which the atoms are transferred from the RDT by adiabatically ramping the power of beam RDT2 down and simultaneously ramping CDT1 up to a power of 300\,mW. The beam RDT1 remains unchanged at a power of 10\,W. The resulting elongated trap geometry with a low atomic density is important to minimize the detrimental effect of re-absorption of scattered photons. After transfer, narrow-line cooling starts with $2.6 \times 10^6$ Dy atoms at a temperature of $\sim 9\,\mu$K.

\begin{figure}[b]
\centering
\includegraphics[width=1\columnwidth]{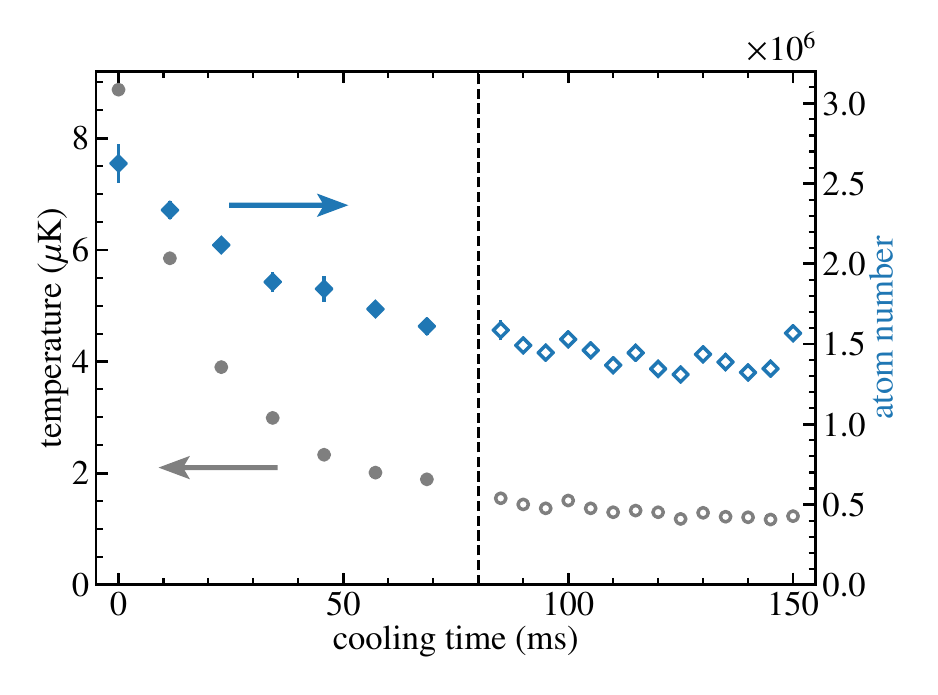}
\caption{Laser cooling using the narrow 741-nm line. The evolution of temperature and atom number is shown within the 150-ms duration of the cooling process. After 80\,ms (vertical dashed line), the laser parameters are changed to optimize the cooling process (see text).}
\label{fig:741cooling}
\end{figure}

In-trap Doppler cooling is carried out in two steps: First, the beam intensity is set to about $6\,\mu$W/cm$^2$. A red laser detuning of 969\,kHz is chosen with respect to the free-space resonance frequency at zero magnetic field, which is larger than light shift introduced by the trap and the Zeeman shift. Figure~\ref{fig:741cooling} shows the evolution of temperature and atom number during the Doppler cooling process. Within 80\,ms (vertical dashed line), the Dy temperature is reduced to 1.7\,$\mu$K, while the number of trapped atoms decreases by roughly a factor of two. We attribute the latter to plain evaporation along the weak axis of the trap. In a second step, we reduce the trap power (RDT1 beam reduced to 1.7\,W within 10\,ms), and we simultaneously decrease the cooling laser detuning and intensity to 532\,kHz and $3\,\mu$W/cm$^2$. Doppler cooling is then continued for 60\,ms, and we finally reach a temperature of $1.2\,\mu$K for $1.6 \times 10^6$ Dy atoms. We found this second Doppler cooling step crucial to accumulate the atoms in the center of the trapping region with maximum overlap with the CDT then used for evaporative cooling.

In a mixture of Dy and K, the K component is not affected by the Dy cooling light and the cooling performance of Dy is not compromised by the presence of K atoms. 

\section{\label{app:BCali}Magnetic field calibration}

\begin{figure}[b]
\includegraphics[width=1\columnwidth]{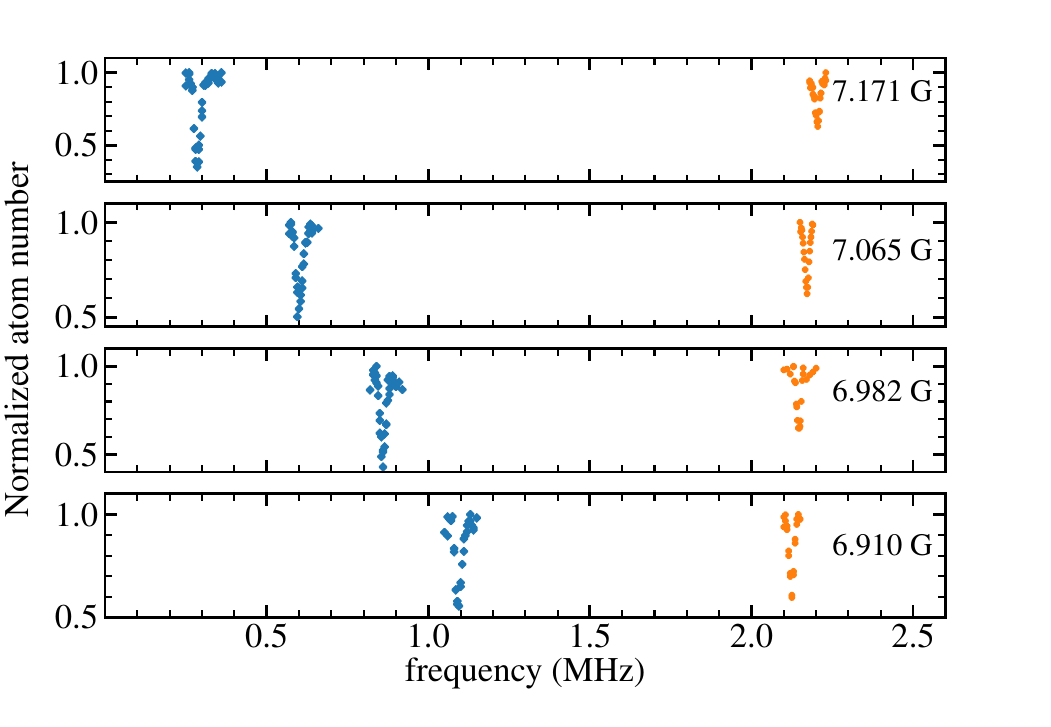}
\caption{\label{fig:BCalibrate}Magnetic field calibration for the binding energy measurements near the 7.29-G resonance. The blue diamond show the signal from molecular association in the Dy-K mixture, and the orange circle show the atomic RF spectroscopic signal in the absence of Dy. The atom number is normalized by the initial values of $N_{\rm K}=3\times10^4$.}
\end{figure}

We calibrate the magnetic field by RF spectroscopy on K. As an example, we here discuss the calibration in the range of interest for the binding energy measurements near the 7.29-G Feshbach resonance. The preparation sequence is the same as described in Sec.~\ref{sec:BEM}. After the ramp to the target field, we blow away all Dy atoms with 421-nm resonant light, leaving only K atoms in the ODT, which populate the lowest hyperfine state K$|1\rangle\,(|F=9/2,m_F=-9/2\rangle)$. We apply a short RF-pulse (100\,$\mu$s) to transfer atoms to the second-to-lowest state K$|2\rangle\,(|F=9/2,m_F=-7/2\rangle)$. Imaging near 7\,G cannot distinguish between the K$|1\rangle$ and the K$|2\rangle$ state, since the Zeeman splitting at 7\,G is much smaller than the linewidth of the transition. To overcome this problem, we ramp up the field strength to a high value (168\,G, limited by the power supply used), where the imaging frequency difference of 28.9\,MHz for imaging K$|1\rangle$ and K$|2\rangle$ is large enough to detect each spin state independently. Time-of-flight imaging is finally applied to record the number of atoms in the state K$|1\rangle$ and K$|2\rangle$.

Figure~\ref{fig:BCalibrate} shows the RF spectroscopic signal on the atomic transition in comparison with the molecular binding energy measurements by magnetic-field wiggle spectroscopy as discussed in Sec.~\ref{sec:BEM}. The signals are compared for four different magnetic field strengths below the 7.29-G resonance. Obviously, the atomic transition (with a differential magnetic moment corresponding to 0.44$\mu_B$) shifts much less than the molecule association signal (shift corresponding to about $-2.2\mu_B$). The magnetic field can be extracted accurately from the atomic signal based on the Breit-Rabi formula. Here, the field uncertainty comes from a fit of a Lorentzian to the signal and typically amounts to about 1.5\,mG.

The calibration of the magnetic field versus current is done in the following way: We first calibrate the magnetic field for different currents, then we extract the dependence of the magnetic field on the current from a linear fit, and finally, the field strengths for other currents are calculated from this linear fit.

We also investigated the magnetic field stability in our setup, again using RF spectroscopy on the K$|1\rangle$\,-\,K$|2\rangle$ transition. The coil setup used for fields up to 1.6\,G possesses a current stability that corresponds to a field stability (the root mean square value of noise) as low as $0.2$\,mG. With the different pair of coils and power supply used to generate higher fields (up to 370\,G), the corresponding stability is 2\,mG. In addition to that, we identify ambient magnetic field noise (mostly 50\,Hz) with a stability of 1.2\,mG. 

\providecommand{\noopsort}[1]{}\providecommand{\singleletter}[1]{#1}%


\begin{thebibliography}{59}
\bibitem [{BCh()}]{BChoice}%
  \BibitemOpen
  \href@noop {} {}\bibinfo {note} {The magnetic field was deliberately chosen
  to minimize the Dy alone loss to reduce the harmful effect on the
  interspecies thermalization process.}\BibitemShut {Stop}%
\bibitem [{Mol()}]{MoltrapDepth}%
  \BibitemOpen
  \href@noop {} {}\bibinfo {note} {The error $\eta$ can be estimated by the
  $(\alpha_M/\alpha_a-1)U_a=\delta\mu\Delta B$, where $U_a$ is the trap depths
  of atoms, $\alpha_M/\alpha_a$ is the polarizability ratio between molecules
  and atoms,$\delta\mu$ is the differential magnetic moment between open
  channel and close channel, such that
  $\xi={(\alpha_M-\alpha_f)}/{\delta\mu\alpha_f}$ and the polarizability of
  molecules can be expressed by
  $\alpha_M=\alpha_f(\xi\delta\mu+1)\approx1.027\alpha_f$.}\BibitemShut {Stop}%
\end{thebibliography}
\end{document}